\documentclass[aps,prl,showpacs,notitlepage,nofootinbib,
superscriptaddress,floatfix,showkeys,twocolumn,
preprintnumbers]{revtex4-1}

\usepackage{blindtext}
\usepackage{hyperref}
\usepackage{amsmath,amssymb}
\usepackage{float}
\usepackage{microtype}
\usepackage{graphicx}
\usepackage{bm}
\usepackage{latexsym}
\usepackage{epsfig}
\usepackage{psfrag}
\usepackage{color}
\usepackage[dvipsnames]{xcolor}
\usepackage{subfigure}
\usepackage[section]{placeins}
\usepackage{comment}

\begin{document}

% ===== MAIN PAPER =====

\title{Sub-keV dark matter can strongly ionize molecular clouds}
%\title{Constraining axion-like dark matter through molecular cloud ionization}

\author{Pedro De la Torre Luque}\email{pedro.delatorre@uam.es}
\affiliation{Departamento de F\'{i}sica Te\'{o}rica, M-15, Universidad Aut\'{o}noma de Madrid, E-28049 Madrid, Spain}
\affiliation{Instituto de F\'{i}sica Te\'{o}rica UAM-CSIC, Universidad Aut\'{o}noma de Madrid, C/ Nicol\'{a}s Cabrera, 13-15, 28049 Madrid, Spain}
%\affiliation{The Oskar Klein Centre, Department of Physics, Stockholm University, Stockholm 106 91, Sweden}
\author{Pierluca Carenza}
\email{pierluca.carenza@fysik.su.se}
\affiliation{Stockholm University and The Oskar Klein Centre for Cosmoparticle Physics, Alba Nova, 10691 Stockholm, Sweden}

\author{Thong T. Q. Nguyen}
\email{thong.nguyen@fysik.su.se}
\affiliation{Stockholm University and The Oskar Klein Centre for Cosmoparticle Physics, Alba Nova, 10691 Stockholm, Sweden}

\smallskip
\begin{abstract}
The ionization of dense molecular clouds offers a powerful and robust probe of dark matter models that produce UV or X-ray photons via annihilation or decay. Using this novel observable, we derive competitive constraints on dark matter in the mass range from 30~eV to 1~keV. In particular, we set the strongest existing limits on axion-like dark matter in the 30–100~eV range. We also present sensitivity projections demonstrating the significant potential of this approach. These constraints can be substantially improved with targeted observations of molecular clouds near the Galactic Center and above the Galactic plane as well as with a more accurate modeling of cosmic-ray ionization and refined gas density estimates from dust extinction maps. Our results establish molecular cloud ionization as one of the most sensitive tools for probing sub-keV dark matter.
 \end{abstract}
\maketitle

\textbf{\emph{Introduction}} ---
Uncovering the fundamental nature of dark matter (DM) remains one of the most pressing challenges in physics. Therefore, it is essential to investigate possible non-gravitational interactions of DM with the visible sector~\cite{Cirelli:2024ssz}.
Light and weakly coupled particle models are attracting significant attention~\cite{Lanfranchi:2020crw,Antel:2023hkf} because of their potential in addressing other open problems, such as  the strong-CP problem. The astrophysical signatures of light DM models are unique, motivating a wide range of indirect detection strategies~\cite{Raffelt:1996wa,Caputo:2024oqc, Jacobsen:2022swa, Foster:2022fxn, Carenza:2024ehj, Nguyen:2025tkl}.

In this work we identify the ionization of dense molecular clouds (MCs) as a new and highly sensitive astrophysical probe of photons injected by light DM annihilation or decay.
The photons injected by sub-keV DM have energies ranging from ultraviolet (UV) to X-ray, and they are able to produce significant ionization of molecular hydrogen, ${\rm H}_2$ in MCs. DM-induced ionization might even exceed the ionization that cosmic rays (CRs) are expected to cause in a standard scenario~\cite{1983ApJ...267..610K,1980ApJ...238..158N}. 
Moreover, DM is a ionization source uniformly distributed across the MC, %, i. e. the DM distribution does not vary significantly on scales of order of a few parsecs,
injecting ionizing radiation directly inside the dense MC. When photons are injected by DM in a dense MC, they are quickly absorbed and efficiently ionize the medium. 
%Notably, a recent study suggests that DM particles with MeV-scale masses might account for the anomalously high ionization rates observed in the Galactic Center~\cite{DelaTorreLuque:2024fcc}.

\begin{figure}[t!]
\includegraphics[width=\linewidth]{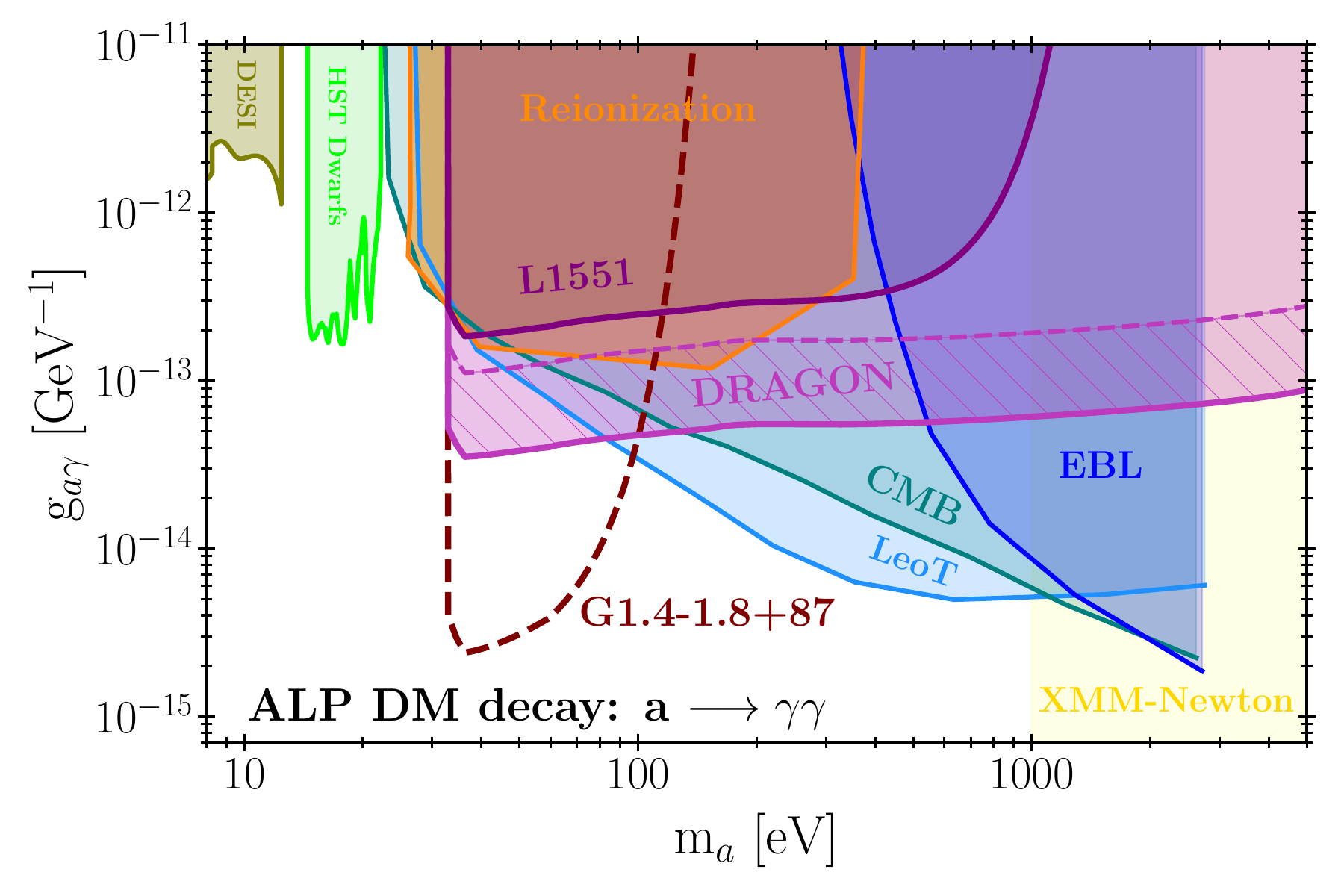}
\caption{Comparison of  previous bounds on ALPs in the $g_{a\gamma}-m_{a}$ plane with the bounds derived in this work. The constraints for two well characterized clouds are shown: {\it L1551} (in purple) and the {\it DRAGON} cloud (in magenta),  where the hatch region denotes its uncertainty. In addition, we show an optimistic forecast based on {\it G1.4-1.8+87} (shown as a brown dashed line to distinguish it from the robust limits). Other constraints come from DESI observations~\cite{Wang:2023imi} (olive), observations of dwarf galaxies by Hubble Space Telescope~\cite{Todarello:2024qci} (HST -- in lime), from reionization~\cite{Cadamuro:2011fd} (orange), CMB spectral distorsions~\cite{Bolliet:2020ofj} (teal), X-ray light~\cite{Essig:2013goa} (blue), heating of gas in the LeoT dwarf at T$_{gas} = 7552$~k~\cite{Wadekar:2019mpc} (dodgerblue) and X-ray observations by XMM-Newton~\cite{Gewering-Peine:2016yoj, Foster:2021ngm} (yellow).}
\label{fig:gag_Lim}
\end{figure}

In the sub-keV mass range, there are many well-motivated extensions of the Standard Model (SM) predicting DM particles able, for example, to generate photons in their decay.
Axion-like particles (ALPs), including the QCD axion~\cite{Peccei:2006as} as a special case, are generically predicted to couple weakly to photons and possibly contribute to DM~\cite{Choi:2020rgn,Gendler:2023kjt}.
The simplest and most motivated QCD axion models~\cite{Peccei:1977hh,Weinberg:1977ma,Wilczek:1977pj} can easily produce cold DM through the misalignment mechanism~\cite{Preskill:1982cy,Abbott:1982af,Dine:1982ah} or topological defect decay~\cite{Harari:1987ht,Hagmann:1990mj,Buschmann:2021sdq,Kawasaki:2014sqa}. 
The robust stellar bounds~\cite{Ayala:2014pea,Carenza:2020zil,Lucente:2022wai} usually preclude the simplest realizations of a QCD action in the eV-keV mass range. However, assuming the existence of $\mathcal{N}\sim\mathcal{O}(10^6)$ degenerate SM copies, with the axion being the common portal in all the replicas~\cite{DiLuzio:2021pxd,Giannotti:2005eb,DiLuzio:2020wdo}, increases the QCD axion mass up to the eV-keV scale. 
Another possibility is that the axion couples to a copy of the QCD sector, which introduces a dynamical energy scale around $\sim 10-100$~GeV~\cite{Rubakov:1997vp}, increasing the axion mass at fixed photon coupling. See also Ref.~\cite{Gavela:2023tzu} for models with many axions and Ref.~\cite{deGiorgi:2024elx} for models with extra dimensions.

The emergence of ALPs in the eV-keV mass range might be unrelated to solving the strong CP problem, and for this reason we refer to them as ALPs and not QCD axions. As a matter of fact, ALPs appear as pseudo-Nambu-Goldstone bosons of a broken global $U(1)$ symmetry. Thus, their mass and couplings will be unrelated. The interactions between ALPs and photons lead to the production of ALPs from the thermal plasma in the early Universe. In the ALP mass range between tens of eV and keV there are stringent constraints requiring the ALP-photon coupling to be $g_{a\gamma}\lesssim 10^{-13}-10^{-14}~{\rm GeV}^{-1}$. For such a small coupling, ALPs will never reach thermal equilibrium with the visible sector, leaving the freeze-in scenario as a possibility. %However, as discussed in Ref.~\cite{Langhoff:2022bij}, it is extremely challenging to produce the totality of DM in form of ALPs. %Moreover, a non-standard thermal evolution might be needed to cool down ALPs in order to be cold DM~\cite{Carenza:2021ebx}. Thus, thermal ALP cold DM is not an appealing possibility.
Non-thermal production mechanisms, such as misalignment and topological defect decay~\cite{OHare:2021zrq,Gelmini:2021yzu}, are a more appealing possibility. The misalignment mechanism, is unlikely to allow for ALP DM in the eV-keV range unless some fine-tuning on the initial conditions is involved. By contrast, depending on the details of the topological defect annihilation, ALPs in the considered mass range might compose the entirety of DM, that leave observable gravitational wave imprints~\cite{Gelmini:2021yzu}. Several QCD axion and ALP models predict viable DM in the eV–keV range, motivating searches in this regime.
%As discussed, there are several models for QCD axion and ALP DM in the eV-keV mass range, making them a plausible DM candidate and motivating searches in this mass range.

In this {\it Letter}, we prove that the MC ionization bound can impose nearly leading constraints on axion DM from 30~eV to 200~eV, as illustrated in Fig.~\ref{fig:gag_Lim}, with extremely good perspectives for future improvements. In Fig.~\ref{fig:gag_Lim}, we compare existent limits with the bounds from different, well characterized, MCs ({\it L1551} and the {\it DRAGON} cloud), and we show an optimistic forecast for a MC near the GC, {\it G1.4-1.8+87}, as described below. This exemplary DM model proves the great potential of using MCs as ultrasensitive natural detectors of DM-induced ionization.

\textbf{\emph{Ionization of dense molecular clouds}} --- 
While MCs with low column densities\footnote{The column density is defined as the integral of the density of a system over our line of sight, $N_{H_2} = \int_{l.o.s.} n_{H_2}{\rm d}l $}, N$_{H_2} \lesssim 10^{22}$~cm$^{-2}$, called diffuse MCs, are expected to be highly affected by external UV radiation penetrating the cloud, above this column density, corresponding to dense MCs~\cite{2006ARA&A..44..367S}, the UV radiation is strongly attenuated, and low-energy CRs are expected to be the dominant ionizing agent~\cite{1989ApJ...345..782M}. In fact, CRs are expected to drive the chemical complexity observed in neutral gas systems and play a major role in the regulation of star formation.
Modeling the contribution of CRs to the ionization of MCs is an intricate problem. Early studies estimated that the ionization rate per atom in clouds composed mainly of atomic hydrogen is within the range of a few$ \times 10^{-18} \ {\rm s}^{-1} \lesssim \zeta^{\rm H} \lesssim 10^{-15} \ {\rm s}^{-1}$~\cite{1968ApJ...152..971S, 2012ApJ...745...91I, Phan:2022iaq, Indriolo_2015}. This broad uncertainty reflects poor knowledge of the CR spectra at energies much below $\sim 10$~ MeV. 
%In particular, the ionization cross section for atomic hydrogen peaks at approximately $50$~eV for CR electrons and around $10$~keV for CR protons~\cite{2009A&A...501..619P}.
%In order to determine with more accuracy the expected ionization rates, a better understanding of the low-energy spectra of CR electrons, protons, and heavier nuclei is required, which is an extremely challenging task~\cite{Padovani:2018ypk, Orlando:2017mvd}. %The simplest approach consists of measuring the low-energy CR spectrum reaching Earth and extrapolating it to energies below the CR observations. Unfortunately, local measurements of the CR spectra at low energies are affected by solar modulation and the structure of the interplanetary magnetic field. The Voyager I and II spacecrafts have provided valuable data, measuring the electron spectrum down to a few MeV~\cite{Cummings:2016pdr,2019NatAs...3.1013S} beyond the heliopause, although these observations are still far from the energy range most relevant for ionization.
%An alternative to probing these low energies involves observations of the Galactic synchrotron background, which is limited %. This method is limited  by significant uncertainties in modeling the interstellar magnetic field~\cite{Padovani:2018ypk, Orlando:2017mvd,Ginzburg:1965su} and it would hardly probe energies of tens of eV without involving a low-energy extrapolation of the CR spectrum. %Thus, modeling the ionization of dense MCs is subject to significant uncertainties.

Experimentally, measuring the ionization of MCs is of interest. A particularly effective method to infer the CR ionization rate involves measuring the intensities of emission lines from molecular tracers, whose abundances are closely tied to the free electron population. The choice of tracer strongly depends on the gas density and chemical composition. In diffuse regions of MCs, ${\rm H}^{+}_{3}$ is a favorable tracer due to its simple and well-understood chemistry~\cite{2013ChSRv..42.7763I}. In denser environments, such as MC cores and proto-stellar clusters, commonly used tracers include HCO$^+$, DCO$^+$, and CO~\cite{1998ApJ...499..234C}. At present, measurements of the ionization rate in MCs seem challenging to explain by CR ionization. 
First, if CRs are the dominant ionization agent of MCs across the Galaxy, one expects that the ionization rate decreases for MCs with larger column densities. Although there are observations that support this trend, no clear correlation is observed~\cite{Gabici:2019jvz, Gabici:2022nac}. Second, the levels of ionization from CRs are expected to be lower than that measured in several clouds~\cite{Indriolo:2009tf, 2018MNRAS.480.5167P}. In this context, the high ionization rate observed in the Central Molecular Zone have been found to be incompatible with what the ionization that CRs can provide~\cite{Ravikularaman:2024umo}, and MeV DM has been proposed as a potential explanation of this anomalous ionization~\cite{DelaTorreLuque:2024fcc}. In addition, there are hints from gamma-ray observations suggesting that CRs do not actually penetrate dense MCs~\cite{Yang:2023vza}.
As opposed to standard ionization mechanisms, UV photon produced by DM interactions would ionize directly within the cloud, providing a constant additional ionization to that provided by CRs. In this work, we explore the potential of dark, or dense, clouds, which are observed with temperatures of tens of Kelvin, sizes from a below the parsec to tens of parsecs and masses up to $\sim10^5$~M$_{\odot}$~\cite{Bergin:2007mt}, to probe the emission of UV/X-ray photons from sub-keV DM.
In order to show how promising are such clouds as targets for DM searches, we will derive constraints from a local dark cloud as the most pessimistic case, because the modeling of the DM density does not need assuming any DM density distribution and will therefore be very robust. Then, we will evaluate constraints from a cloud closer to the Galactic Center (GC), where DM is expected to be concentrated, and we will assume a prototypical Navarro-Frenk-White (NFW)~\cite{Navarro:1995iw} distribution. %This is the standard assumption made in most of the indirect DM searches, and allows for a fair estimation with previous constraints. 
%As we will explain below, we follow a very conservative approach in both cases, and better results can be obtained through a more dedicated search of optimal clouds that is left for future work.

One of the local clouds that has received more attention is {\it L1551}, which is located $150$~pc away from the Earth in the direction of the Taurus constellation and has a size of $\sim 1$~pc. This cloud presents a low ionization rate, with $\zeta^{H_2} \approx (5 \pm 1.5) \times 10^{-18} \,\, \textbf{s}^{-1}$~\cite{deBoisanger:1995je}. We will derive constraints at $95\%$ confidence level by requiring the DM-induced ionization to match the observed ionization at the $2 \sigma$ level. Moreover, we note that we are not considering here the standard ionization that would be produced by CRs or UV/X-rays, which makes our estimations extremely conservative. We will discuss the expected improvement when CR ionization is modeled.

We stress that {\it L1551} is not the local cloud with the lowest ionization rate. For example, the Taurus molecular cloud (TMC-1) is a dark cloud in the constellations of Taurus and Auriga. For this cloud, Ref.~\cite{2007A&A...474..923F} found that $\zeta^{H_2} \approx 2 \times 10^{-18}\,\, \textbf{s}^{-1}$ provided the best overall agreement with the observations. Using this cloud would improve our limits by a factor $2-3$ on the DM annihilation/decay rate. 
However, the lack of a robust estimation of the uncertainties of this measurement makes us choose {\it L1551} as reference case.
Another interesting example is {\it Barnard 68}, which was found to be compatible with an ionization rate of $(1-6) \times 10^{-18} \,\, \textrm{s}^{-1}$~\cite{2007ApJ...664..956M}. 
Unfortunately, there is no precise estimation of the uncertainty of these measurements that allow us to use them reliably, but it must be noted that using these clouds would lead to at least a factor of a few stronger DM constraints than {\it L1551}.
Lower values of ionization can be found in the local environment, but we leave a dedicated survey for a future work. 

Interestingly, similar or even lower values of ionization have been found in infrared dark clouds (IRDCs), which are very cold (T$<20$~K) and dense clouds, and can be ideal targets for DM searches. Among known IRDCs, one that has been very well studied recently is {\it G28.37+00.07}, often referred to as the {\it DRAGON} cloud, which is located around $5$~kpc away from the GC (in the so-called $5$-kpc ring)~\cite{2006ApJ...653.1325S}. This cloud has a size around L$\sim13.6$~pc, a density of n$_{H_2}\approx 987$~cm$^{- 3}$ and a column density N$_{H_2} \approx 2.1\times10^{22}$~cm$^{- 2}$. 
Ref.~\cite{2022A&A...662A..39E} studied the ionization rate and other gas properties across different sub-regions of this cloud, finding that the ionization rate in some sub-regions of the cloud was exceptionally low. Indeed, the authors of this paper chose some of these sub-regions because they were expected to gather no star formation activity. Since the ionization from photons produced by DM is expected to be uniform throughout the cloud, due to negligible variations in DM density on scales of a few parsecs, constraints on DM-induced ionization can be derived from measurements in individual zones within the cloud.
In particular the region P8 of that study was found to have ionization values between $10^{-19}$~s$^{-1}$ and $4.6\times10^{-19}$~s$^{-1}$ in their fiducial model. We set an upper limit on the ionization of this cloud by taking twice the highest value found in this analysis, i.e. $\zeta_{lim} = 9.2\times10^{-19}$~s$^{-1}$, which corresponds to our fiducial constraint. The authors of the same paper also performed an additional analysis that included more ionized species, noting that these species have more uncertain astrochemistry and are less directly linked to the induced ionization rates, which may introduce additional uncertainty into their analysis. This analysis yielded ionization rates of $10^{-19},\mathrm{s^{-1}}$ and $4.6\times10^{-18},\mathrm{s^{-1}}$. Accordingly, we present the results for the {\it DRAGON} cloud as a band, adopting constraints in the range $9.2\times10^{-19},\mathrm{s^{-1}}$ to $9.2\times10^{-18},\mathrm{s^{-1}}$
The P8 region has a size of $0.8$~pc, with a gas density around $10^6$~cm$^{-3}$.

While the modeling of DM ionization in local clouds would not require to assume any DM distribution, we also select a cloud located quite close to the GC to demonstrate the potential of such dark clouds to probe sub-keV DM. The advantage is the higher DM density in this region, with the disadvantage of introducing uncertainties related to the DM distribution profile. For concreteness, we choose {\it G1.4-1.8+87}, with a size of around $8.2$~pc, and located around $400$~pc away from GC, above the Galactic plane~\cite{Bhoonah:2018wmw}. This cloud was found to be compatible with an ionization rate as low as $1.9 \times 10^{-19 }\,\, \textrm{s}^{-1}$ % and $5 \times 10^{-18}\,\,  \textrm{s}^{-1}$
~\cite{Bhoonah:2018gjb}, however, Ref.~\cite{Farrar:2019qrv} casted serious doubts on the gas temperature used in this analysis. We will use this cloud as an example of the optimistic case of MC with low ionization rate located close to the GC. However, this should not be interpreted as a strict constraint, but rather as a forecast to illustrate the potential of similar clouds once more robust analyses are performed. To derive this limit,  we choose a value of ionization of 
$3.8 \times 10^{-19}\,\, \textrm{s}^{-1}$ and will assume a prototypical NFW DM profile, which is the standard assumption in this kind of searches. We will call this an optimistic constraint, that will be shown in a different way than the fiducial and conservative cases, for clarity.
As in the case of local MCs, more research is needed to identify more clouds close to the GC.  %IRDCs located close to the GC but above the Galactic plane are probably the best target to constraint the ionization caused by sub-keV DM, since the impact of CR ionization will be significantly reduced.

\textbf{\emph{Ionization of UV/X-ray photons from keV DM}} --- 
The ionization rate $\zeta_{H_2}$, for a MC located at $\textbf{x}$ caused by photons with energy $E$, can be calculated as~\cite{Padovani:2009bj,Phan:2022iaq}
\begin{equation}
  \zeta_{H_2} =  \int_{I_{H_2}}^{\infty} J_{\chi}(E, \textbf{x}) \, \sigma^{\text{Ioniz}}_{H_2}(E)\,(1 + \theta_e(E)) \, dE 
   \,\,,
\label{eq:main}
\end{equation}
where $J(E, \textbf{x})$ is the photon flux in a given location of the sky, $\sigma^{\text{Ioniz}}_{H_2}(E)$ is the photoionization cross section~\cite{Padovani:2009bj} and $\theta_e(E)$ is the number of secondary ionizations per primary ionization~\cite{2001ApJ...559.1194Y, Dogiel:2013mwa}
\begin{equation}
\theta_e (E_e) = (E_e - I_{H_2})/W \, .
\label{eq:SecIoniz}
\end{equation}
Here, $I_{H_2} = 15.4$~eV is the ionization potential of H$_2$, $E_e = E -I_{H_2}$ is the energy of the electron produced from the photoionization induced by an incoming photon with energy $E$ and $W \lesssim  40$~eV is the average energy loss of a non-thermal electron per ionization of an H$_2$ molecule in the interstellar medium~\cite{1999ApJS..125..237D}. Precisely, the values of W we taken from Fig. 6 of Ref.~\cite{1999ApJS..125..237D}.
The integral in Eq.~\ref{eq:main} extends from a minimum energy of $I_{H_2}$ to the maximum photon energy produced in the DM annihilation/decay, i.e. the mass of the DM particle, m$_{\chi}$, in the case of annihilating DM and m$_{\chi}$/2 in the case of decaying DM. Here, we are neglecting the Compton ionization process, which is important only above tens of keV~\cite{2001ApJ...559.1194Y}. This simplification is motivated by the fact that at such energies, our limits are not competitive with previous constraints.

The photon flux at a given position $J(E,\textbf{x})$ can be expressed as
\begin{equation}
    J_{\chi}(E, \textbf{x}) = 2\cdot Q(E, \textbf{x}) \frac{N_{H_2}}{n_{H_2}} 
     \,,
\label{eq:flux}
\end{equation}
where the factor $2$ accounts for the production of two photons, $\frac{N_H{_2}}{n_{H_2}} \sim$ L$_{MC}$ is of the size of the MCs (L$_{MC}$)~\cite{Gabici:2022nac} calculated for a constant average n$_{H_2}$ and $Q$ is the source term for the injection of photons
\begin{equation}
    Q (E, \textbf{x}) = 
    \left\{
\begin{array}{ll}
	\displaystyle  
        \frac{\langle \sigma v\rangle}{2} \left(\frac{\rho_\chi(\textbf{x})}{m_\chi}\right)^2\frac{dN^{\textrm{ann}}}{dE} &\text{(annihilation)}\\
	\\
        \displaystyle
        \ \; \Gamma \ \; \left(\frac{\rho_\chi(\textbf{x})}{m_\chi}\right)\frac{dN^{\textrm{dec}}}{dE} &\text{(decay)}
    \end{array}
\right.
.
\label{eq:source}
\end{equation}
Here, $\langle \sigma v\rangle$ and $\Gamma$ are the DM thermally averaged annihilation cross section and the DM decay rate, respectively. The DM energy density at the position $\textbf{x}$ is $\rho_\chi({\bf x})$ and $dN/dE$ is the $\gamma$-ray yield per DM interaction, that is a Dirac delta peaking at m$_{\chi}$, in the case of annihilation, or at m$_{\chi}$/2 in the case of decay.

For the local cloud {\it L1551}, we set the DM density to be the local DM density, which is $\rho_{\chi}^{\odot} \sim 0.4$~GeV/cm$^3$~\cite{Benito:2020lgu} at the Earth position $\sim 8.277$~kpc from the GC~\cite{Evans:2018bqy}. We stress that this makes our estimation independent on the DM density profile, which is one of the main uncertainties in DM indirect searches. %In fact, given that the considered cloud is located in the Solar System neighborhood, this assumption leads to a quite conservative result. 
For the case of the {\it DRAGON} cloud, which corresponds to our fiducial constraint, we assume the NFW profile~\cite{Navarro:1995iw}, as it is the most usual assumption used for DM searches. %Considering constraints on the DM lifetime $\tau$, using the NFW profile leads to a factor  $\sim4$ stronger bound than assuming the local DM density at the cloud position, while for annihilation ($\langle\sigma v\rangle$) the difference is close to a factor $10$.  
%We stress that cuspier DM profiles, like the ones favored by the GC excess~\cite{Fermi-LAT:2017opo}, are commonly used as optimistic scenarios, but we will not consider them.
We also adopt a NFW profile for our optimistic forecast with {\it G1.4-1.8+87}.
We note that the local DM density value that we have taken is in the lower bound given in Ref.~\cite{Benito:2020lgu}, where the estimated local DM density is $0.4-0.7$~GeV/cm$^3$ at 2$\sigma$ level.

The previous estimate assumes complete absorption of the UV photons emitted by DM within the cloud; however, this assumption does not always hold. We thus compute the opacity of the cloud ($\tau$) to photons emitted by DM. The resulting absorption probability, $P_{\rm abs} = 1 - e^{-\tau}$, is shown in Fig.~S2 of the Supp. Material as a function of photon energy. In our calculations, specifically in Eq.~\ref{eq:main}, we multiply the integrand $J_\chi$ by this absorption probability to account for photons that escape the cloud, which suppresses our constraints towards high photon energies.

\begin{figure*}[th!]
\includegraphics[width=0.49\linewidth]{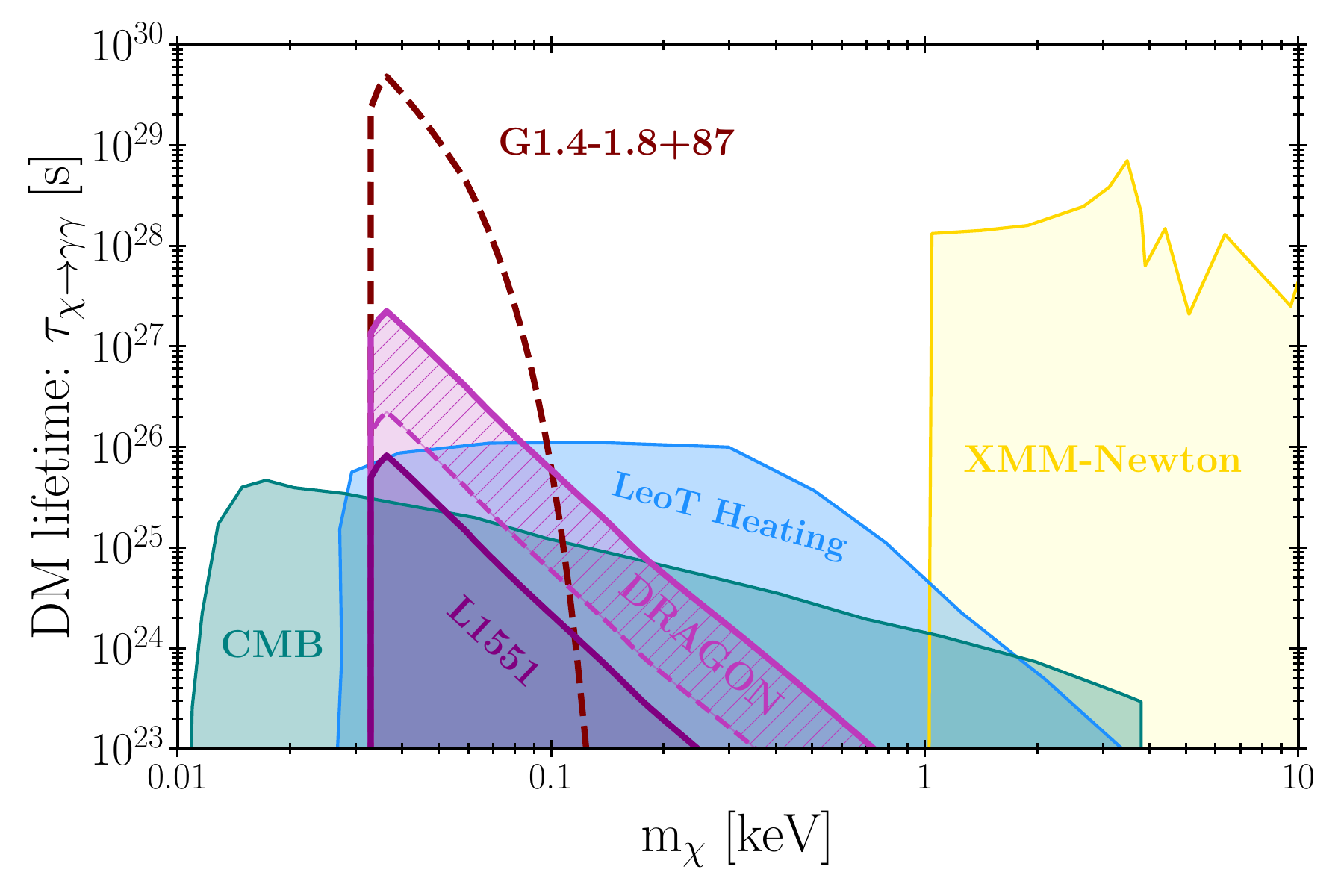}
\includegraphics[width=0.49\linewidth]{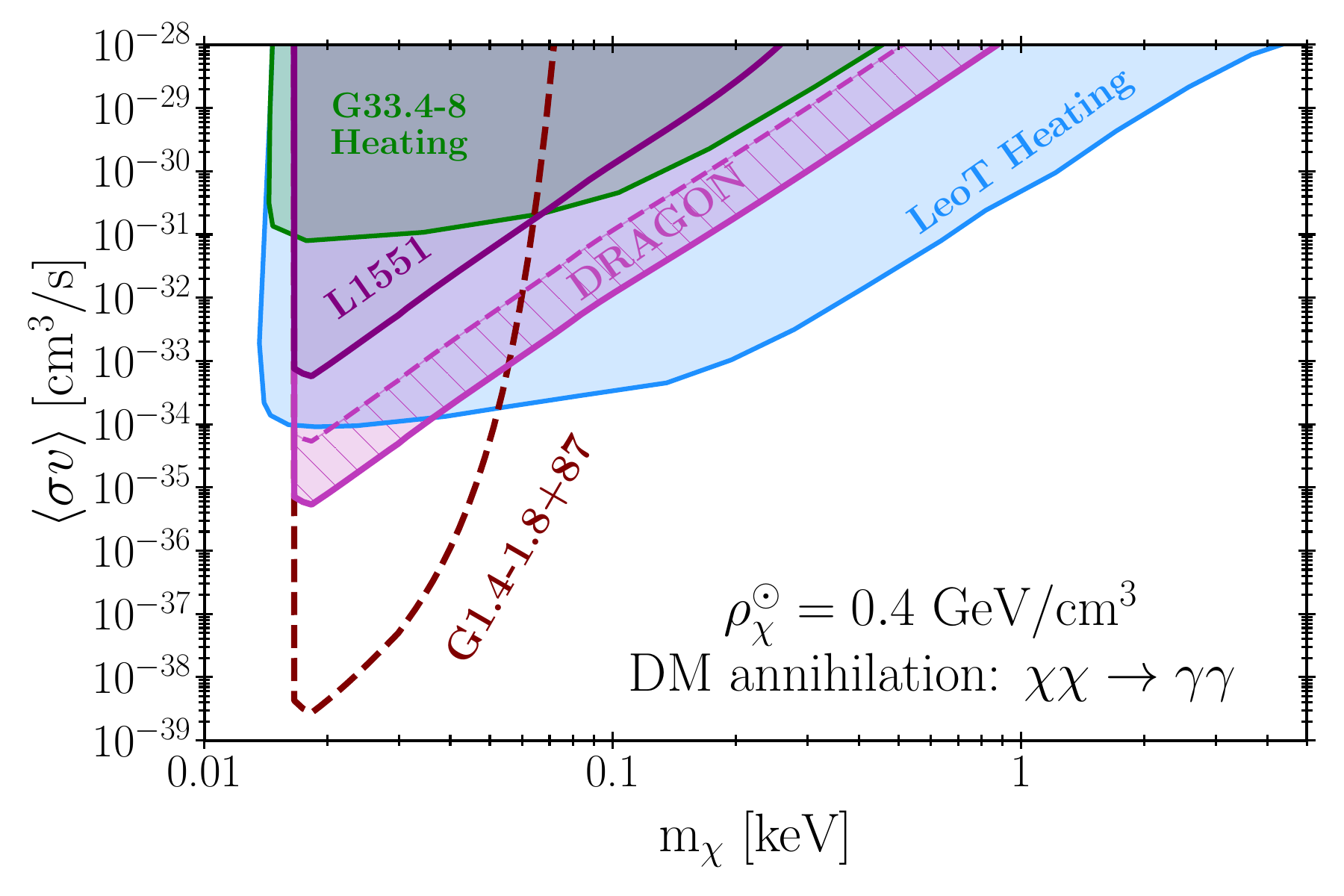}
\caption{Model-independent constraints on the lifetime (left panel) and annihilation rate (right panel) for DM particles producing two photons. We show our constraints from {\it L1551} in purple, from the {\it DRAGON} cloud in magenta and the optimistic forecast based on {\it G1.4-1.8+87} as a brown dashed line. We compare with previous results from CMB spectral distortions (teal)~\cite{Bolliet:2020ofj}, LeoT (dodgerblue) and {\it G33.4-8} (green) gas heating at T$_{gas} = 7552$~K~\cite{Wadekar:2019mpc}, and XMM-Newton diffuse Galactic observations (yellow)~\cite{Essig:2013goa, Gewering-Peine:2016yoj, Boyarsky:2018tvu, Foster:2021ngm}.}
\label{fig:Ioniz_Lims}
\end{figure*} 
\textbf{\emph{Results}} ---
In Fig.~\ref{fig:Ioniz_Lims}, we report our model-independent constraints on DM two-photon decay (left panel) and annihilation (right panel). In this energy range, the strongest constraints are from CMB spectral distortions~\cite{Bolliet:2020ofj} and heating of gas-reached dwarf galaxies (particularly, the Leo T dwarf)~\cite{Wadekar:2021qae}. %The latter constraints not only requires the adoption of a model for the DM distribution, gas temperatures and the density distributions of molecular and atomic gas phases as well as dust grains (which are uncertainties that can similarly affect the measurement of the ionization rate of the {\it DRAGON} cloud), but it also requires modeling the background heating from CRs and UV/X-ray. Thus, different assumptions imply different uncertainties that are involved in these constraints. 
Here, we show that even in the most pessimistic case of a local MC such as {\it L1551}, with a low DM density, the constraints at tens of eV are competitive with the existent bounds~\cite{Wadekar:2021qae, Bolliet:2020ofj, Audren:2014bca, Poulin:2016nat}. Notably, our fiducial bound (the {\it DRAGON} cloud), becomes stronger than CMB constraints in the mass range from $\sim30$~eV to several hundreds eV, and it is also stronger than the LeoT gas heating constraints up to a hundred eV. As commented above, the assumption of a NFW profile for the {\it DRAGON} cloud makes our constraint only a factor of $\sim4$ stronger than assuming only the local DM density at the cloud position, and around a factor of $10$ for the annihilation rate constraint. Taking a Burkert profile~\cite{Burkert:1995yz}, which is often taken as the most pessimistic scenario in Galactic DM searches, these limits are expected to be reduced by a factor $\sim 2$ for decay and $\sim5$ for annihilation (see Fig.~S2 in section S.I. of the Supp. Material, where the parameters characterizing the DM density distribution used are given). We remind the reader, that even cuspier profiles than NFW, like the ones favored by the GC excess~\cite{Fermi-LAT:2017opo} are equally likely, and those would result in an improvement of this limit. On top of this, we note that the presence of DM minihalos or clumps within these clouds will substantially improve the predicted ionization rates, therefore leading to much stronger constraints. We remain conservative by only considering the subjacent DM distribution.
%We stress that cuspier DM profiles, like the ones favored by the GC excess~\cite{Fermi-LAT:2017opo}, are commonly used as optimistic scenarios, but we will not consider them.
Notably, for the optimistic case with {\it G1.4-1.8+87} we get a constraint that is far stronger than previous observations below a hundred eV above the keV. We tested another interesting candidate, the cloud {\it G357.8-4.7-55}, which is around $1.1$~kpc~\cite{Bhoonah:2018wmw} far from the GC and is also compatible with low ionization rates. Assuming NFW profile, we get a lifetime constraint one order of magnitude stronger than the  {\it L1551} MC for decaying DM.

We have also estimated what could be the effect of adding the ionization from CRs in {\it L1551}. Since CRs are expected to be the ionization agents in the interstellar medium, their effect will be dominant. In fact, previous predictions expect the ionization of {\it L1551} to be totally explained by CRs~\cite{Phan:2022iaq}. We adopt the CR ionization calculated in Ref.~\cite{Phan:2022iaq}, pointing towards an ionization rate between $10^{-17}$~s$^{-1}$ and $5.1 \times 10^{-18}$~s$^{-1}$ in the diffusive regime. %This will be the contribution to the ionization of {\it L1551} explained by standard physics. 
We adopt the most conservative value of this estimation, $\zeta_{H_{2}} = 10^{-17}$~s$^{-1}$  in our calculation, and find that adding this contribution leads to around a factor of $3$ better constraints below $1$~keV for both decay and annihilation. This rough approach illustrates the improvement expected in the derived limits when accounting for the background ionization of CRs. 
A similar estimation could be made for clouds that are located farther away from the Solar System, but this would require a precise modeling of the distribution of low-energy CRs, and the uncertainties would be very important.

These model-independent constraints for DM producing two photons can be directly expressed as constraints in the coupling of different feebly interacting particles, such as ALPs, scalars or sterile neutrinos. We calculate our constraint on the coupling of ALPs with photons as~\cite{Carenza:2024ehj}
\begin{equation}
    g_{a\gamma} = \sqrt{\frac{64\,\pi}{m_a^3\,\tau_{\chi} }}
     \,,
\label{eq:gagg}
\end{equation}
where $\tau_{\chi}$ is the model independent limit on the DM lifetime and m$_a$ is the ALP mass.

In Fig.~\ref{fig:gag_Lim} we report bounds for the ALPs-photon coupling, g$_{a\gamma}$, for keV mass ALPs, compared to previous constraints. Even in the most pessimistic case of {\it L1551}, these constraints are competitive with constraints from reionization~\cite{Cadamuro:2011fd} and extend to higher masses. %where they are comparable  with CMB limits and those from the heating of gas in Leo T, which is the leading constraint in the $30$~eV-$1$~keV range. 
For our fiducial constraint, from the {\it DRAGON} cloud, we significantly improve on the CMB and LeoT constraints up to several hundreds of eV.
%We also point out that constraints based on observations of Leo T are influenced by different sources of uncertainty in their model of the gas distribution, their temperature and its DM halo~\cite{Wadekar:2021qae}, that are expected to be similar to those from the {\it DRAGON} cloud. %The halo model used to derive the constraint is discussed in Ref.~\cite{Faerman:2013pmm}, whose uncertainties can mildly affect the resulting constraint. However, a more important issue is connected to the possible breaking of the hydrostatic equilibrium assumption, fundamental for the model of Ref.~\cite{Faerman:2013pmm}. Indeed, even though observations do not point to a rotation of the HI gas in Leo T, possible that the gas is rotating in a direction aligned with our view, which would conceal its motion. 
%However, Ref.~\cite{Wadekar:2021qae} claimed that these uncertainties are under control and the failure of hydrostatic equilibrium might even lead to a stronger bound. Nevertheless, it is evident that our proposed MC bound relies on more controllable assumptions.
Moreover, our optimistic forecast for {\it G1.4-1.8+87}, that represents what we expect from a low ionization cloud near the GC, illustrates that MC observations could be the a significant constraints in the sub-keV range for ALPs and similar particles. We also provide constraints for other well-motivated DM models in the Supplementary Material. This Supp. Material contains the following references:~\cite{2001ApJ...559.1194Y, Burkert:1995yz, Moore_1999, Cirelli:2010xx, Di_Mauro_2021, Dev:2020jkh, Nguyen:2024kwy, 2021MNRAS.507.3148B, Wadekar:2019mpc, Audren:2014bca, Poulin:2016nat, Gewering-Peine:2016yoj, Essig:2013goa, Foster:2021ngm, Slatyer:2015jla, Fiorillo:2025zzx, Bottaro:2023gep, Bolliet:2020ofj, Boyarsky:2018tvu, Wadekar:2021qae, Bezrukov:2009th, Bringmann:2022aim, Tremaine:1979we, Dev:2025sah, Pospelov:2008jk, Linden:2024uph, McDermott:2017qcg, Cheung:2025gdn, Nguyen:2025eva, Redondo:2008ec, Krnjaic:2022wor, Linden:2024fby, Holdom:1985ag, Fayet:1980ad, Fayet:1980rr, Kahn:2016vjr, Fabbrichesi:2020wbt, Caputo:2021eaa, Chun:2022qcg, Carena:2004xs, Nguyen:2022zwb, Smith:2024jve, Servant:2002aq, Graham:2015rva, Hebecker:2023qwl, Cyncynates:2023zwj, Cyncynates:2024yxm, Carenza:2025uwx, XENON:2019gfn}

\textbf{\emph{Discussion and conclusions}} ---
In this {\it Letter}, we investigate the ionization of neutral gas produced in dense MCs by decaying/annihilating DM. In particular, sub-keV DM could generate UV/X-ray photons that are able to produce a high and homogeneous ionization across the cloud since they are absorbed efficiently within such systems. In fact, we show that the ionization produced by the photon emission from sub-keV DM saturates the ionization that is observed in different MCs for decay lifetimes, or annihilation rates, that are not ruled out by any other constraint. Therefore, we test for the first time MCs as a target for indirect DM searches, demonstrating that they allow us to improve current constraints for decaying DM particles such as ALPs, dark photons, scalar particles or sterile neutrinos for masses from a few tens of eV to slightly below $1$~keV.

%Interestingly, The ionization of dense MCs is expected to be primarily driven by CRs, as UV light cannot effectively penetrate these regions. However, sub-keV unstable DM particles could be producing UV radiation directly inside the MC, leading to a persistent background of ionization, potentially influencing both the star formation rates and the chemical evolution within the clouds.
%Notably, a recent study suggests that DM particles with MeV-scale masses might account for the anomalously high ionization rates observed in the Galactic Center~\cite{DelaTorreLuque:2024fcc}.

We begin by identifying local MCs as a particularly robust probe of this kind of DM. This is because estimating their ionization rate requires only knowledge of the local DM density  without the need to assume a specific DM distribution model. Using the local MC {\it L1551} we derive constraints that are similar, or even better, than the previous leading constraints. 
%Many of the existing bounds rely on assumptions that can affect their reliability and expose them to significant uncertainties, as opposed to the more robust MC bounds that we explored.
Then, we examine the case of {\it G28.37+00.07}, or the {\it DRAGON} cloud, which has been studied previously in detail and shows regions of it that are exceptionally quiescent in terms of their star formation activity, with ionization rates compatible with $10^{-19}$~s$^{-1}$. This constitutes what we call our fiducial constraint, and allows us improve previous constraints from CMB and LeoT gas heating up to several hundreds of eV.
Finally, we use the case of {\it  G1.4-1.8+87}, an MC $400$~pc away from the GC, to show the potential of using clouds near the GC. For this case, we take the values obtained in Ref.~\cite{Bhoonah:2018gjb} to illustrate that, assuming an NFW DM density profile, we could reach the strongest DM bounds to date in a mass range from a few tens of eV to the keV. However, this must be taken as an optimistic forecast since the characterization of the temperature of that could was put in doubt in Ref.~\cite{Farrar:2019qrv}. Clouds that are closer to the GC can be found, therefore providing even stronger DM bounds.
Finally, we highlight that the most promising targets for indirectly searching for sub-keV DM decays or annihilations into photons are IRDCs located near the GC but slightly above the Galactic plane.  These regions are expected to be less affected by CR ionization, which is sourced in the Galactic disk. As a result, such clouds are likely to exhibit very low ionization levels, making any DM-induced effects more distinguishable. Although a more comprehensive survey of MCs will further refine the constraints presented here, our results already demonstrate the strong potential of this approach for probing different types of light DM particles. %We also note that this strategy can be also applied for MCs in nearby galaxies.

However, even though these results are quite promising, one must keep in mind that the uncertainties in the measured ionization rates of Galactic molecular clouds are significant. These analyses require a combination of observations of molecular abundances and column densities to inform chemical models and infer the underlying ionization rates. Several factors contribute to these uncertainties. The chemistry of molecular clouds is highly sensitive to local conditions such as density, temperature, and radiation fields, which can vary across the cloud and are not always precisely known. Observational limitations, including instrumental sensitivity, line-of-sight integration, and assumptions about cloud geometry, further add to the uncertainty. Additionally, many studies report only best-fit values without fully quantified statistical errors, so the derived ionization rates often represent estimates rather than precise measurements. Despite these limitations, these analyses remain the most direct method to probe ionization processes in molecular clouds, and careful modeling allows researchers to account for uncertainties by considering conservative scenarios or quoting results as ranges. 
In this line, the James Webb Space Telescope has performed the first direct measurement of the ionization rate of Barnard 68~\cite{bialy2025detectioncosmicrayexcitedh2}, which opens a window for refined ionization rate measurements in the next years.

We list a few expected future refinements below: \newline
i) A more exhaustive survey of the measurements of dark MCs, dedicated to identify clouds with lower ionization fractions and closer to the GC. %These measurements are often difficult, and they are  usually focused on measuring the ionization around clouds with regions of star formation. However, 
Such dedicated exploration and analysis could yield improvements of up to a few orders of magnitude in the derived DM constraints.
We show that dark clouds constitute a very interesting target for DM searches, hoping to increase the interest towards targeting dark MCs for observations. %In addition, if dark clouds systems are closer to the GC, these limits can be highly improved.
%In this regard, there are already extensive surveys of such clouds, like the analyses from Ref.~\cite{2006ApJ...653.1325S}. 
 \newline
ii) A recent paper~\cite{2024ApJ...973..142O} showed that previous measurements of the ionization rate from several dense MCs are biased, due to an inaccurate evaluation of their density. Using dust extinction maps to evaluate their density leads to an ionization rate that is lower by an average factor of $9$. Analyzing interesting targets using this procedure could lead to constraints that are an order of magnitude better than the current ones. \newline
iii) Probably the most obvious improvement would be that of modeling the ionization produced by CRs. Provided that CRs are expected to be the main ionization agent of dense MCs, including their contribution must certainly improve our constraints. Unfortunately, this estimation is not simple and requires an accurate knowledge of the distribution of CRs in different regions of the Galaxy and at very low energies, where our models are not robust enough to produce a reliable constraint. Reasonable estimates for local MCs already yield order-of-few improvements under conservative assumptions, as discussed above.
 %However, reasonable estimations can be made for local MCs, and improvements of a factor of a few in the constraints can be already obtained with conservative assumptions, as discussed above.

%As a final point, we also note that X-ray telescopes are very promising for constraining DM above a few hundreds eV. Especially eROSITA could place constraints above $0.2 $~keV that can beat all previous constraints~\cite{Dekker:2021bos}, once they have enough observational time. However, the the current constraints from eROSITA or Chandra below $1$~keV are even weaker than those that we obtain and forecast in this work.

In conclusion, we have shown the potential impact of sub-keV DM decaying/annihilating into photons in dense gas regions, which can be specially relevant for probing ALPs and other DM particles. Our results demonstrate that astrochemical observations in MCs can serve as a novel and powerful probe for indirect searches of sub-keV DM. This approach offers particularly robust constraints across a range of particle physics models and has the potential to surpass existing bounds for DM masses between approximately $30$~eV and $1$~keV.

\newpage
\textbf{\emph{Acknowledgements}} --- 
We thank Torsten Bringmann for fruitful discussion and Rebecca Leane for her useful comments and feedback. PDL has been supported by the Juan de la Cierva JDC2022-048916-I grant, funded by MCIU/AEI/10.13039/501100011033 European Union "NextGenerationEU"/PRTR, and is currently supported by Ramón y Cajal RYC2024-048445-I grant, which is funded by MCIU/AEI/10.13039/501100011033 and FSE+. The work of PDL is also supported by the grants PID2021-125331NB-I00 and CEX2020-001007-S, both funded by MCIN/AEI/10.13039/501100011033 and by ``ERDF A way of making Europe''. PDL also acknowledges the MultiDark Network, ref. RED2022-134411-T. This project used computing resources from the National Academic Infrastructure for Supercomputing in Sweden (NAISS) under project NAISS 2024/5-666.
PC and TTQN are supported by the Swedish Research Council under contract 2022-04283. This article/publication is based on the work from COST Action COSMIC WISPers CA21106, supported by COST (European Cooperation in Science and Technology).

%\newpage

\bibliographystyle{apsrev4-1}
\bibliography{references.bib}

%%%%%%%%%%%%%%%%%%%%%%%%%%%%%%%%%%%%%%%%%%%%%%%%%%%%%%%%%%%%%%%%%%%%%%%%%%%%%%%%

\clearpage
\onecolumngrid   % switch to 1 column for supplement (revtex trick)

% ===== SUPPLEMENTARY MATERIAL =====
\section*{Supplementary Material}

% redefine numbering
\setcounter{figure}{0}
\renewcommand{\thefigure}{S\arabic{figure}}
\renewcommand{\thesection}{S\Roman{section}}

\title{Sub-keV dark matter can strongly ionize molecular clouds}
%\title{Constraining axion-like dark matter through molecular cloud ionization}

\author{Pedro De la Torre Luque}\email{pedro.delatorre@uam.es}
\affiliation{Departamento de F\'{i}sica Te\'{o}rica, M-15, Universidad Aut\'{o}noma de Madrid, E-28049 Madrid, Spain}
\affiliation{Instituto de F\'{i}sica Te\'{o}rica UAM-CSIC, Universidad Aut\'{o}noma de Madrid, C/ Nicol\'{a}s Cabrera, 13-15, 28049 Madrid, Spain}
%\affiliation{The Oskar Klein Centre, Department of Physics, Stockholm University, Stockholm 106 91, Sweden}
\author{Pierluca Carenza}
\email{pierluca.carenza@fysik.su.se}
\affiliation{Stockholm University and The Oskar Klein Centre for Cosmoparticle Physics, Alba Nova, 10691 Stockholm, Sweden}

\author{Thong T. Q. Nguyen}
\email{thong.nguyen@fysik.su.se}
\affiliation{Stockholm University and The Oskar Klein Centre for Cosmoparticle Physics, Alba Nova, 10691 Stockholm, Sweden}

\bibliographystyle{apsrev4-1}

\maketitle

%%%%%%%%%%%%%%%%%%%%%%%%%%%%%%%%%%%%%%%%%%%%%%%%%%%%%%%%%%%%%%%%%%%%%%%%%%%%%%%%

\section{Additional details}

\begin{figure}[hb!]
\includegraphics[width=0.49\textwidth]{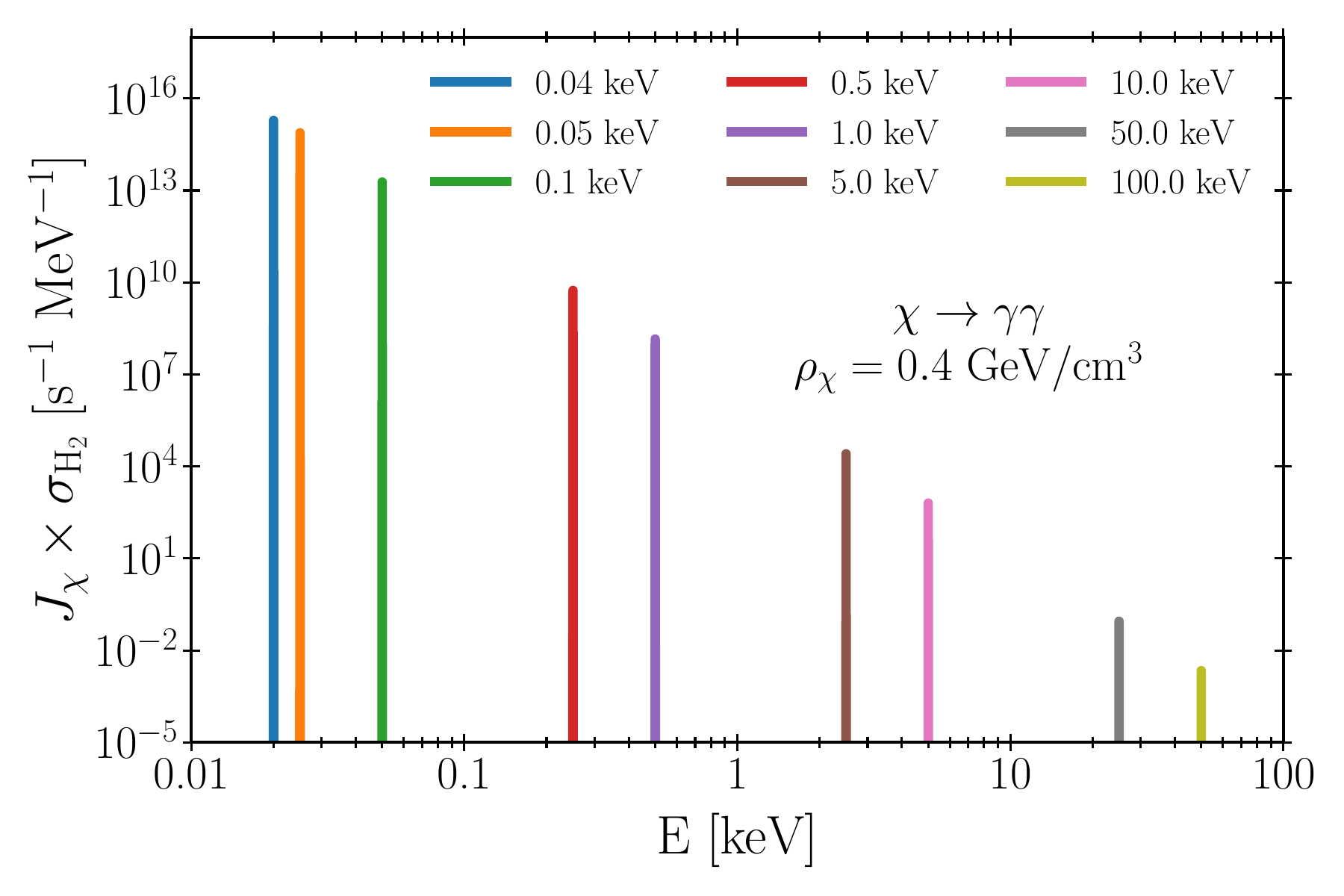}
\includegraphics[width=0.49\textwidth]{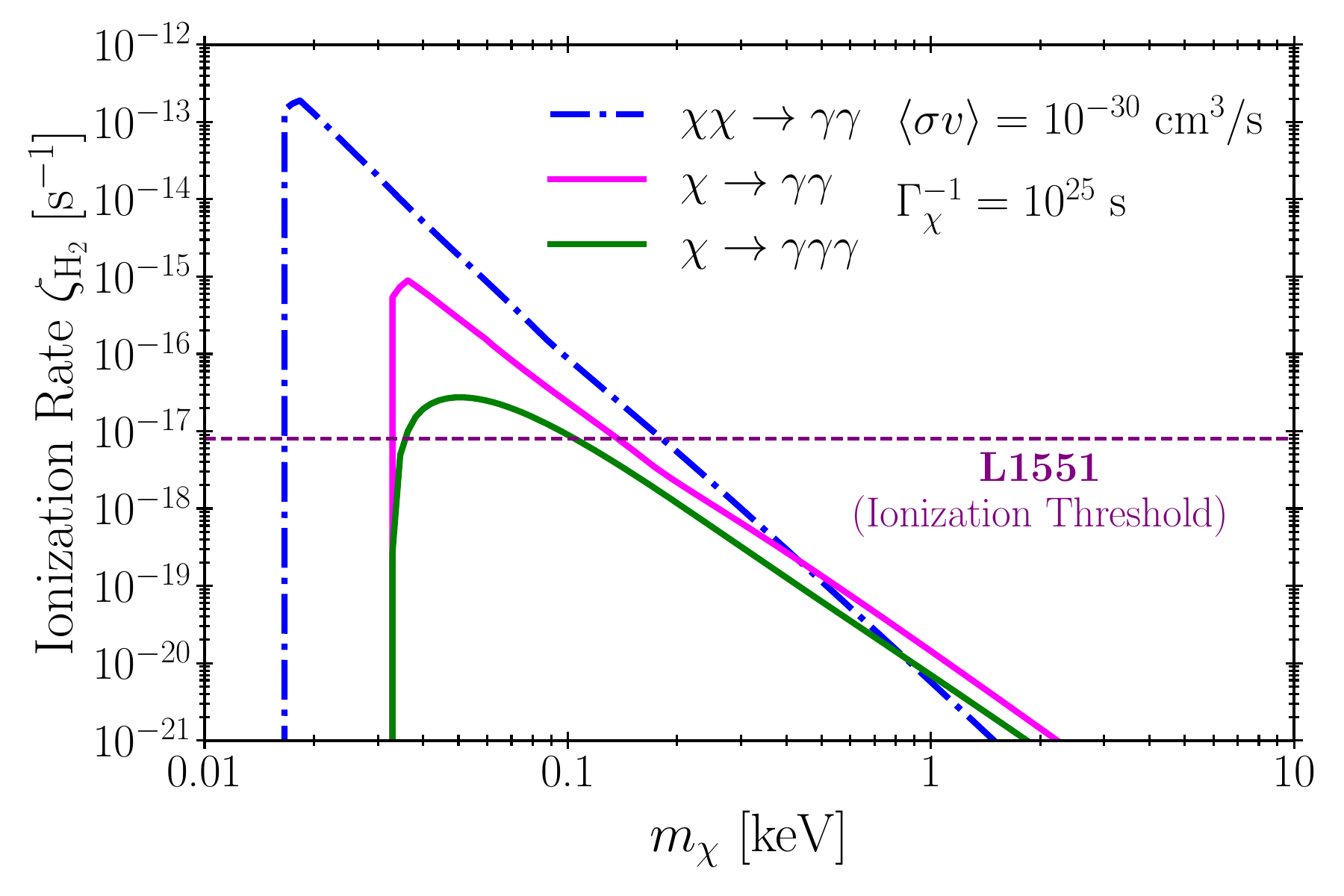}
\caption{\textbf{left panel:} Product of DM-induced photon flux times ionization cross section, J$_{\chi}  \sigma_{H_2}$, for the decay into two photons. This serves as a direct indicator of the ionization rate, for several DM masses, and assumes $\rho_{\chi} =0.4$~GeV/cm$^{3}$, modeling a local MC like {\it L1551}. \textbf{Right panel:} Ionization rate as a function of the DM mass, induced by two- and three-body decays (solid lines) with a rate $\Gamma=10^{-25}$~s$^{-1}$ or annihilation (dot-dashed line) for an averaged cross section $\langle\sigma v\rangle=10^{-30}~{\rm cm}^{3}~{\rm s}^{-1}$. These calculations refer to the MC {\it L1551}, whose $2\sigma$ upper limit in ionization rate is shown by the purple dashed line. }
\label{fig:Ioniz}
\end{figure}

In this section, we provide additional figures that allow us to describe better our calculations.
In the left panel of Fig.~\ref{fig:Ioniz} we show the quantity J$_{\chi}  \sigma_{H_2}$ for several DM masses in a local MC like {\it L1551}. As expected, the lower the DM mass, the higher is J$_{\chi}  \sigma_{H_2}$ and, thus, the ionization rate. This behavior is due to the higher DM number density and the higher photoionization cross sections, which peaks around $20$~eV~\cite{2001ApJ...559.1194Y}. This can be also seen from the right panel of Fig.~\ref{fig:Ioniz}, where we show the H$_2$ ionization rate caused by annihilating (dot-dashed line) and decaying (solid lines) DM in a local MC. The purple dashed line shows the level of ionization observed for {\it L1551}.
We note that in this work we will constrain the annihilation rate into photons defined as $\langle \sigma v \rangle_{\gamma } = \text{BR}_{\gamma } \times \left< \sigma v \right>_{\rm ann}$, where $\left< \sigma v \right>_{\rm ann}$ is the total DM annihilation rate and $\text{BR}_{\gamma }$ is the branching ratio for photons in the final state. Similarly, we will always refer to DM decay into photons.

\begin{figure}[ht!]
\includegraphics[width=0.6\textwidth]{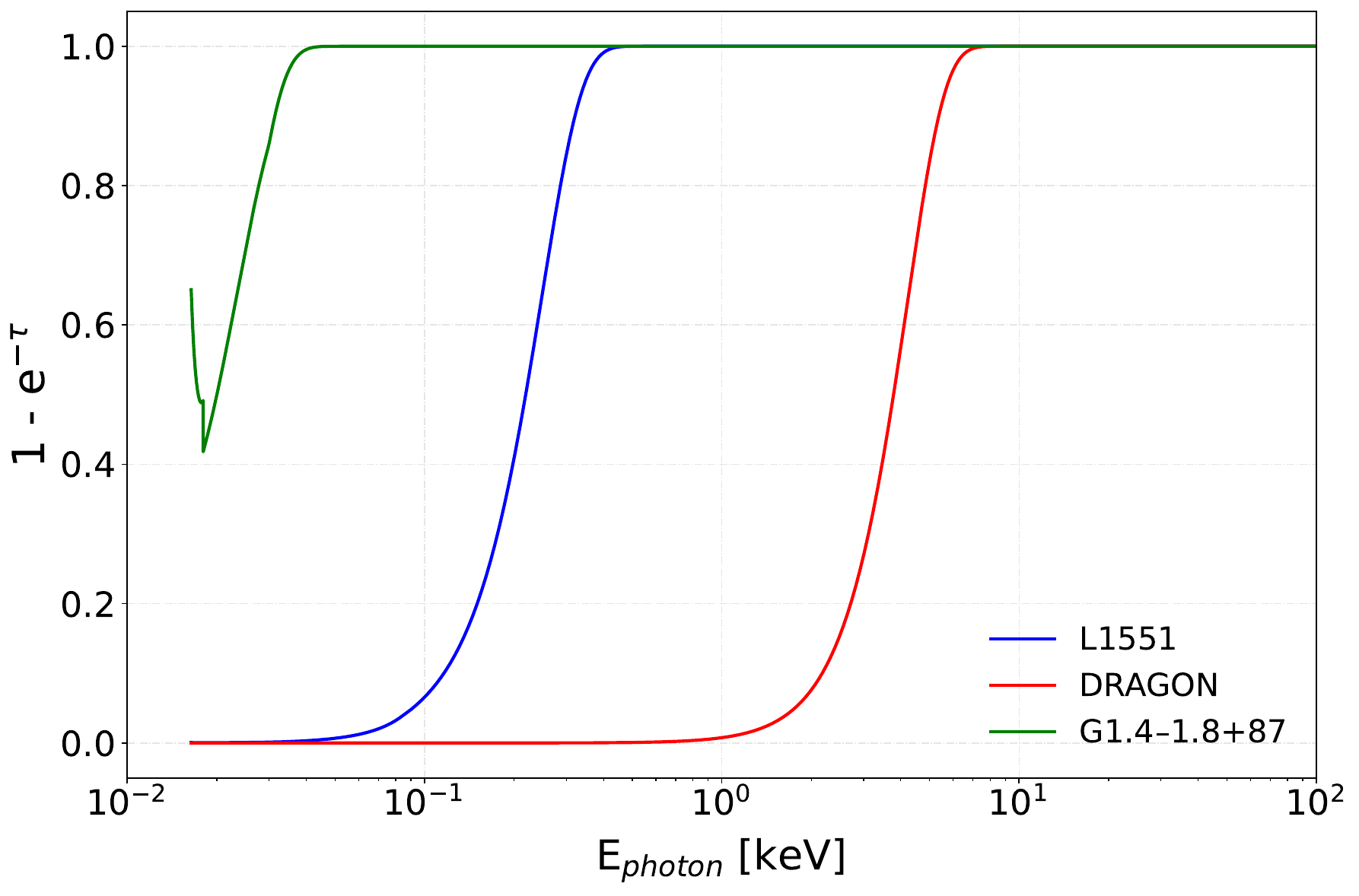}
\caption{Absorption probability of photons emitted by DM as a function of their energy, for each of the clouds studied here.}
\label{fig:AbsProb}
\end{figure}

Another important factor in our study is the opacity of each cloud, which characterizes how efficiently photons can escape from the MC. The opacity is computed as $\tau (E) = \frac{L_{MC}}{n_H \, \sigma^{\text{Ioniz}}_{H_2}(E)}$.
A very high opacity (i.e. an absorption probability close to unity) implies that photons cannot escape the molecular cloud and instead deposit all of their energy into the cloud in the form of ionization. As we see from Fig~\ref{fig:AbsProb}, while the high density of the P8 region in the {\it DRAGON} cloud makes it effectively opaque to photons with energies below 1 keV, in the {\it L1151} and {\it G1.4–1.8+87} clouds there exists an energy above which photons escape efficiently, producing little ionization above a few hundred keV.

\begin{figure}[ht!]
\includegraphics[width=0.49\textwidth]{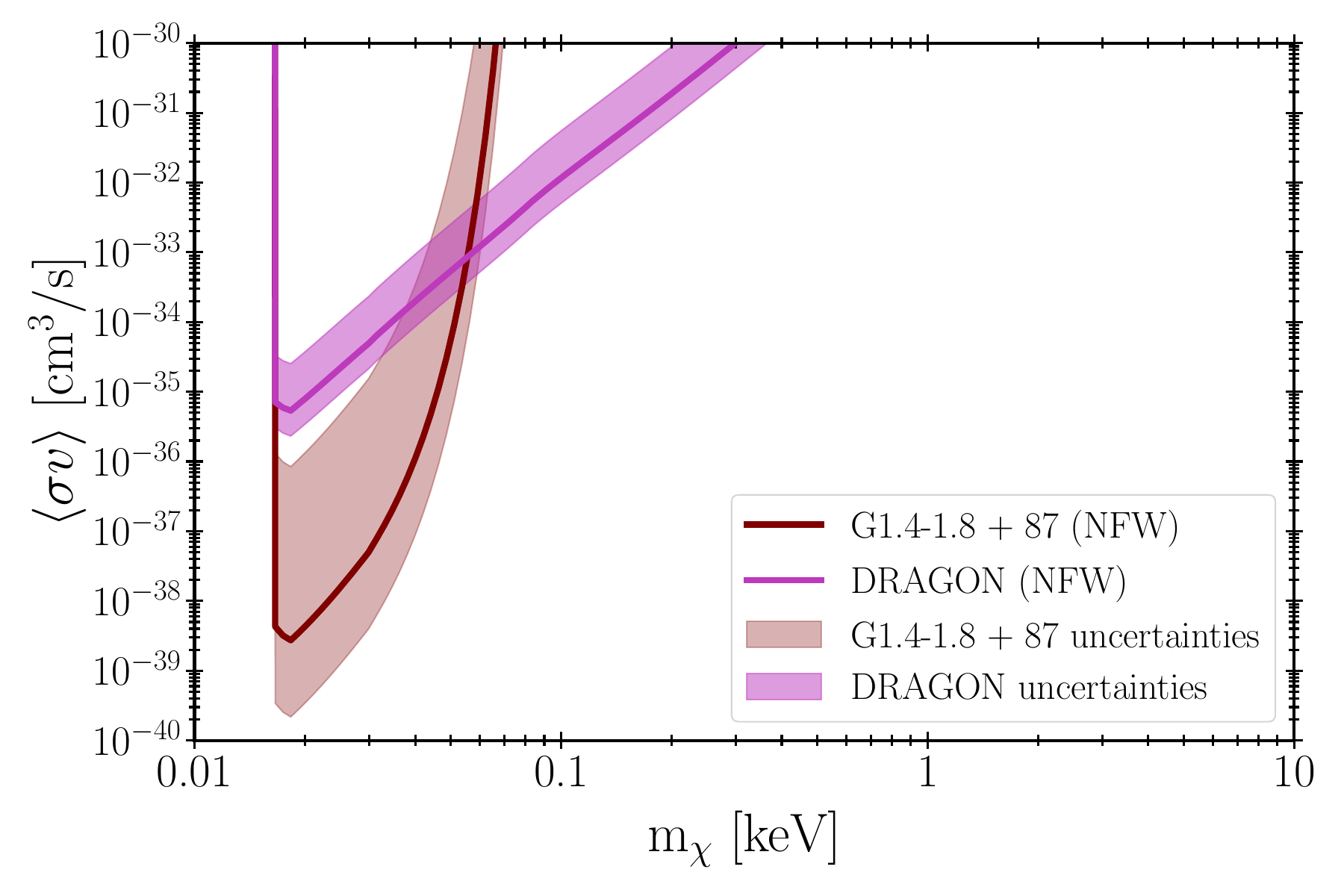}
\includegraphics[width=0.49\textwidth]{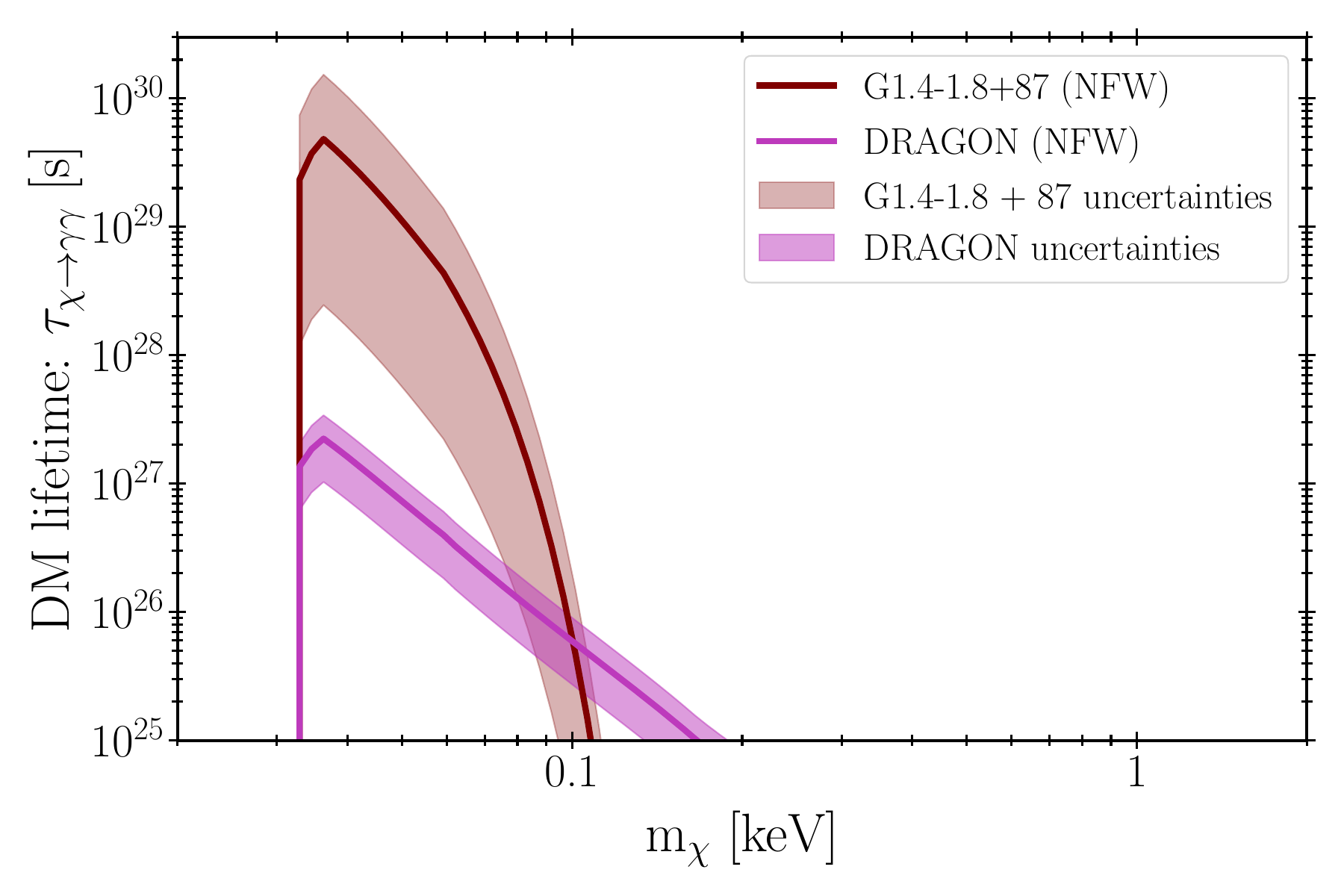}
\caption{Constraints on  annihilation rate $\langle \sigma v \rangle$ (left panel) and lifetime $\tau$ (right panel), for the {\it DRAGON} cloud (in magenta) and the forecast for {\it G1.4-1.8+87} (in maroon). In every case, the bands indicate the difference between the constraint obtained when assuming a Burkert and a Moore profile, and the solid lines represent the constraint obtained assumed an NFW profile.}
\label{fig:Uncerts}
\end{figure}

\newpage
Then, we show the uncertainties in our model-independent constraints on $\tau$ and $\langle \sigma v \rangle$ for the {\it DRAGON} and {\it G1.4-1.8+87} clouds associated to the assumed DM density profile. In particular, we take two extreme assumptions for the DM profile: the most pessimistic choice, which is the Burkert profile~\cite{Burkert:1995yz} (Eq.~\ref{eq:BurkProf}) and the most optimistic choice, a Moore-like profile~\cite{Moore_1999} (characterized by a contraction index $\gamma=1.5$) (Eq.~\ref{eq:MooreProf}). Their parametric forms are the following: 
\begin{equation}
\rho_{\mathrm{NFW}}(r) = \frac{\rho_0}{\frac{r}{r_s}\left(1 + \frac{r}{r_s}\right)^2}\,\,.
\end{equation}
\begin{align}
    \rho_{\mathrm{Moore}}(r) = \frac{\rho_0}{\left(\frac{r}{r_s}\right)^{1.5}\left(1 + \frac{r}{r_s}\right)^{1.5}} \,\,.
\label{eq:MooreProf}
\end{align}
\begin{equation}
\rho_{\mathrm{Burkert}}(r) = \frac{\rho_0}{\frac{r}{r_s}\left(1 + \frac{r}{r_s}\right)^2}\,\,.
\label{eq:BurkProf}
\end{equation}
For the Burkert profile (Eq.~\ref{eq:BurkProf}), we use a scale radius of $R_s = 12.67$~kpc~\cite{Cirelli:2010xx}, while for the Moore profile, we employ $R_s = 28$~kpc. For the NFW profile, that is used as reference, $R_s = 20$~kpc~\cite{Di_Mauro_2021}. In all these cases the density normalization, $\rho_0$ is found by imposing that the local DM density is $\rho (r=r_{\odot}) = 0.4$~GeV/cm$^3$. As we see in Fig.~\ref{fig:Uncerts}, the constraint obtained from the ionization in the {\it DRAGON} cloud is not very affected by the choice of DM density distribution, either for decay of annihilation, which makes this cloud an optimal target for robust DM searches. However, the projected constraint based on {\it G1.4-1.8+87} varies significantly depending on the choice of DM distribution, making it less robust, but with tremendous potential under the assumption of a cuspy DM density distribution in the Galaxy.

\section{Other particle models}
\begin{figure}[t!]
\includegraphics[width=0.49\linewidth]{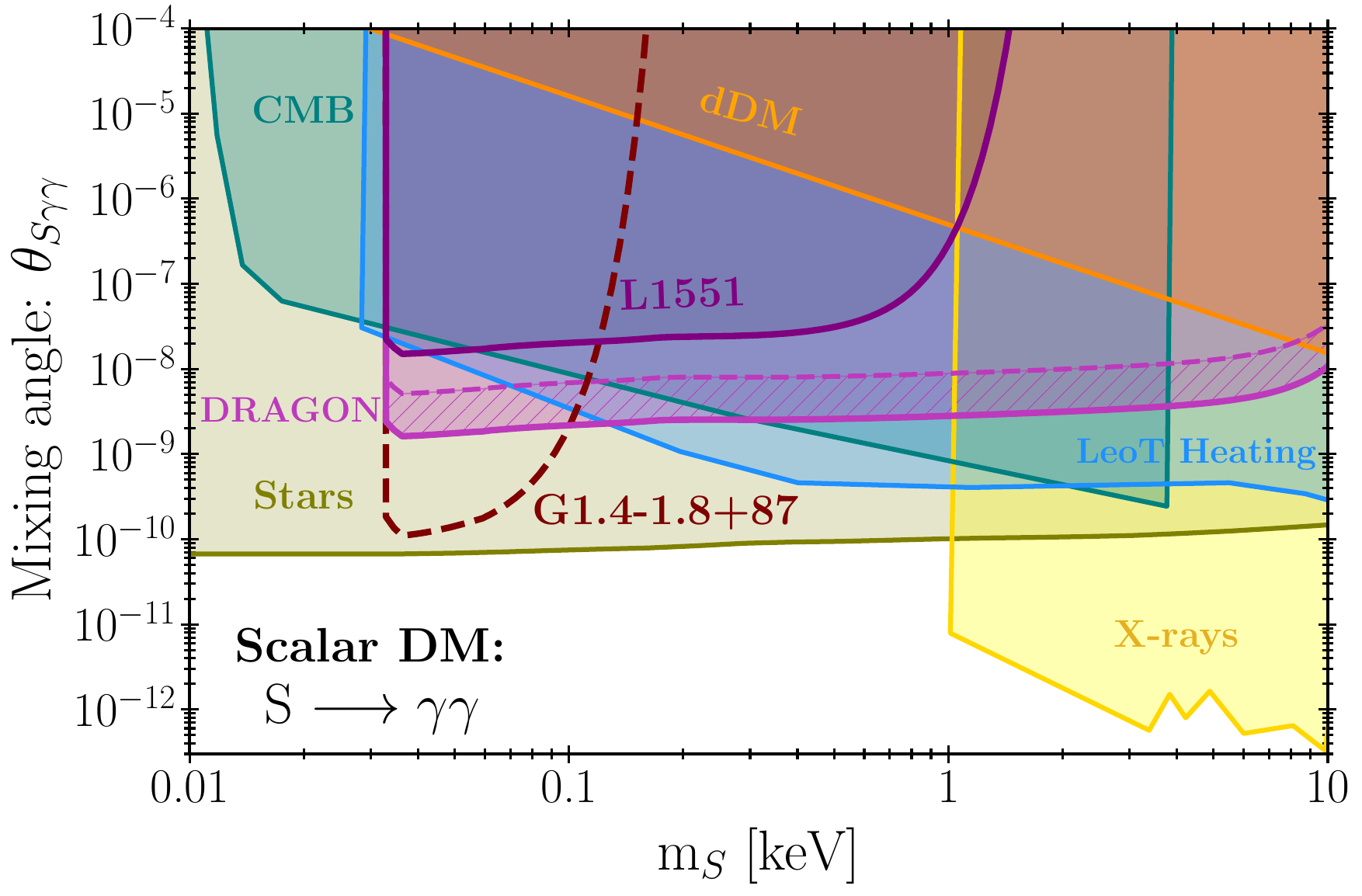}
\includegraphics[width=0.49\linewidth]{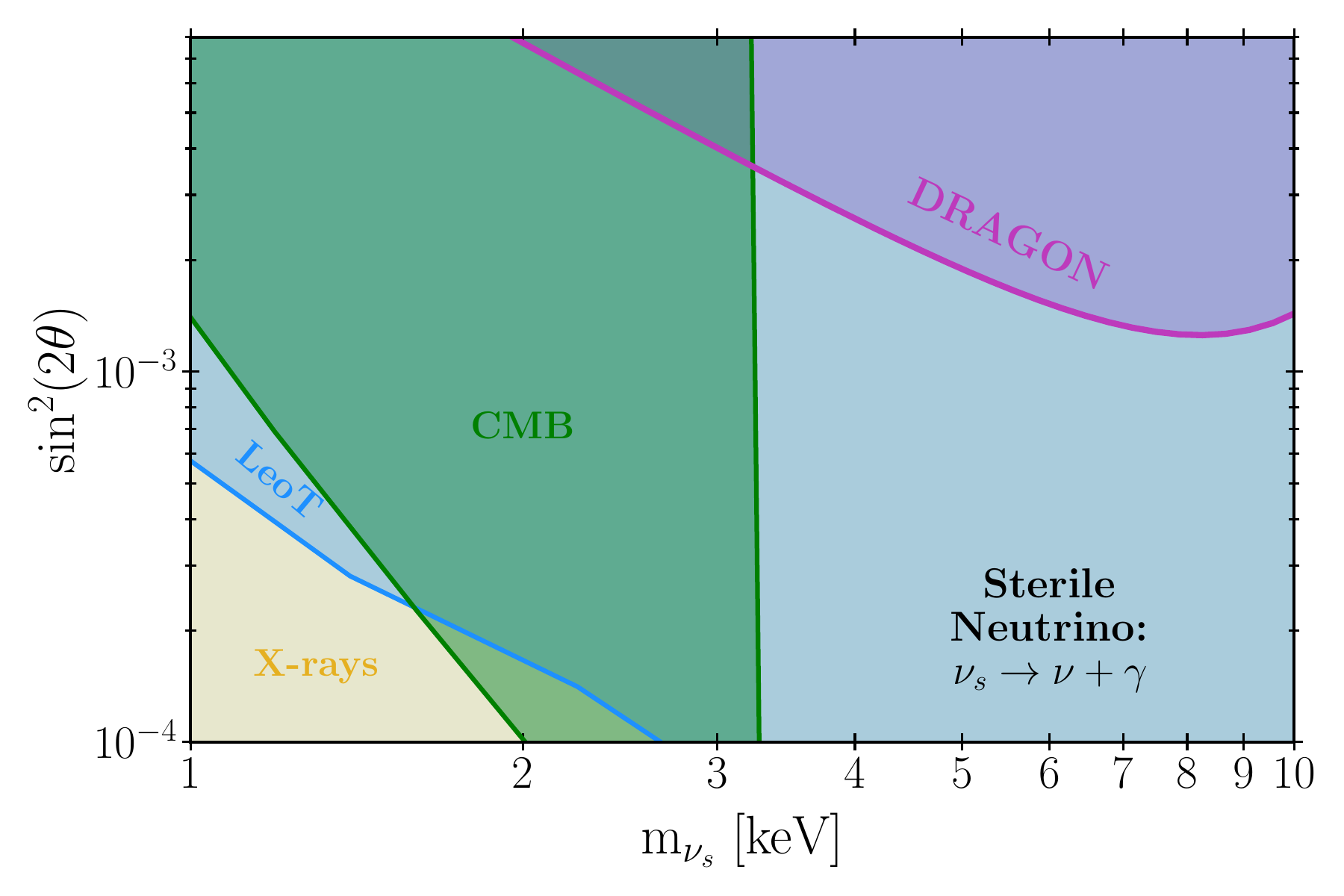}
\caption{\textbf{Left panel:} Compilation of current bounds on scalar dark matter, compared with our limits derived from the local cloud \emph{L1551} (purple), the \emph{DRAGON} cloud (magenta), and the optimistic forecast from \emph{G1.4–1.8+87} (dashed maroon). Shaded regions indicate constraints from CMB spectral distortions (teal)~\cite{Bolliet:2020ofj}, X-ray observations (yellow)~\cite{Essig:2013goa, Gewering-Peine:2016yoj, Boyarsky:2018tvu, Foster:2021ngm}, general decaying dark matter (orange)~\cite{Audren:2014bca, Poulin:2016nat}, Leo T gas heating (dodger blue)~\cite{Wadekar:2021qae}, and stellar luminosity limits (olive)~\cite{Fiorillo:2025zzx}. \textbf{Right panel:} Comparison of the current constraints for the sterile neutrino mixing angle (in the form of sin$^2(2\theta)$), with the DRAGON constraint derived in this work. Constraints from CMB spectral distortion are shown in teal, those from X-ray observations in yellow, the Lyman-alpha bound is shown in grey, those from Leo T gas heating in dodgerblue.} %and the Tremaine-Gunn limit is shown as a light-green band. }
\label{fig:S_Lim}
\end{figure}

While the case of ALPs is probably the most interesting among current light DM candidates, in this appendix we explore the constraints for other particle models coupling to photons.
First, we report the constraint for scalar particles as DM, that have mixing angle $\theta_{S\gamma\gamma}$ with the SM Higgs~\cite{Dev:2020jkh, Nguyen:2024kwy}. We calculate this mixing angle from the scalar dark matter decay to two photons lifetime as %we report the constraint for scalar particles as DM, where their coupling constant $\theta$ follows the following expression~\cite{Dev:2020jkh}
\begin{equation}
    \theta_{S\gamma\gamma} = \sqrt{\frac{9}{121}\frac{512\,\pi^2 \, v_{\rm EW}^2}{\alpha^2 \, m_S^3\,\tau_{\chi} }}
     \,,
\label{eq:Sgg}
\end{equation}
where $\alpha \approx \frac{1}{137}$ is the fine structure constant, $v_{\rm EW}=246$~GeV is the electroweak vacuum expectation value and m$_S$ is the mass of the scalar particle.
A comparison of our constraints with previous ones, for scalar DM, is shown in the left panel of Fig.~\ref{fig:S_Lim}. As for the case of ALPs, we are able to improve on existent constraints, such as those from CMB spectral distortions~\cite{2021MNRAS.507.3148B}, LeoT gas heating~\cite{Wadekar:2019mpc} or the model independent constraint on decaying DM (dDM) derived in Refs.~\cite{Audren:2014bca, Poulin:2016nat}, and our results are complementary with X-ray constraints~\cite{Gewering-Peine:2016yoj, Essig:2013goa, Foster:2021ngm}, constraints from the CMB anisotropies~\cite{Slatyer:2015jla} and stellar constraints~\cite{Fiorillo:2025zzx,Bottaro:2023gep}. Interestingly, MCs can probe a region of the parameter space that cannot be currently tested from arguments of the stars dynamics.

We also compute our limit for sterile neutrinos, which have a decay of the form $\nu_S \rightarrow \nu\, \gamma$. The expression used for our calculation is~\cite{Boyarsky:2018tvu, Bezrukov:2009th}
\begin{equation}
   {\rm sin}^2(2\theta) = \frac{1024\,\pi^4}{9\, \alpha \, G_F^2 \, m_{\nu_S}^5\,\tau_{\chi} }% \approx 4 \theta^2 \,\,\text{(for low $\theta$ values)}
     \,,
\label{eq:tnunug}
\end{equation}
where m$_{\nu_S}$ is the sterile neutrino mass and G$_F$ is the Fermi constant that represents the strength of the weak force. In this case, instead of producing two photons in the decay, only one is produced. Therefore, the $\tau_{\chi}$ value adopted is a factor of 2 lower than the two-photon final state.
%Here, we specify the limit for low $\theta$ values, since we show our constraints for sterile neutrinos in the $\theta^2$ vs mass plane. T
These bounds are shown in the right panel of Fig.~\ref{fig:S_Lim}, compared to previous ones. As one can see, the parameter space that is covered by MCs is already well tested by other mechanisms, specially the Lyman-alpha forest~\cite{Bringmann:2022aim}. Also the Tremaine-Gunn constraint~\cite{Tremaine:1979we} rules out sterile neutrinos with masses below $\sim0.5$~keV. We report the optimistic constraint derived from \emph{G1.4-1.8+87}, which may become relevant again if model-dependent approaches reopen this region of parameter space~\cite{Dev:2025sah}.

\begin{figure}[t!]
\includegraphics[width=0.49\linewidth]{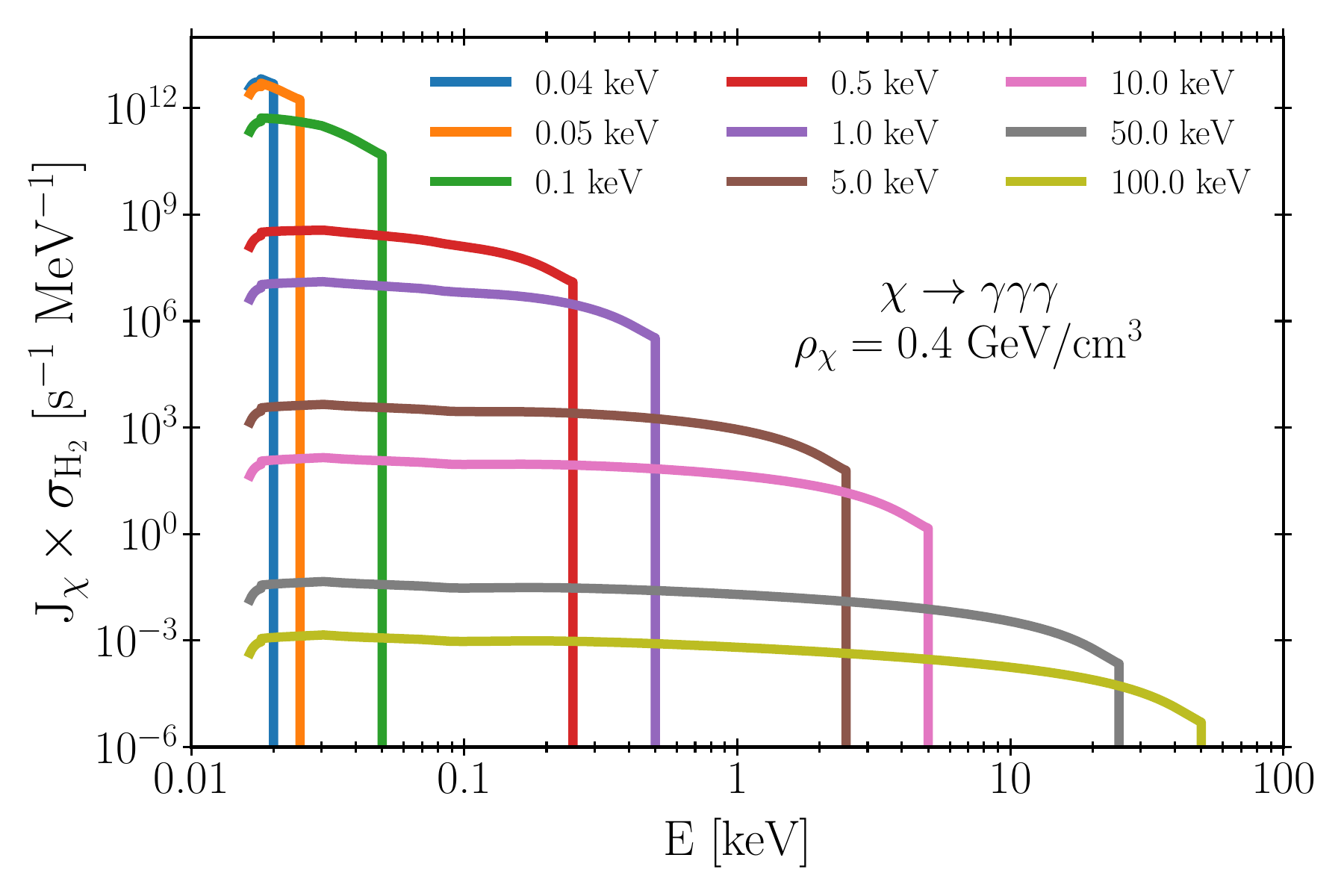}
\includegraphics[width=0.49\linewidth]{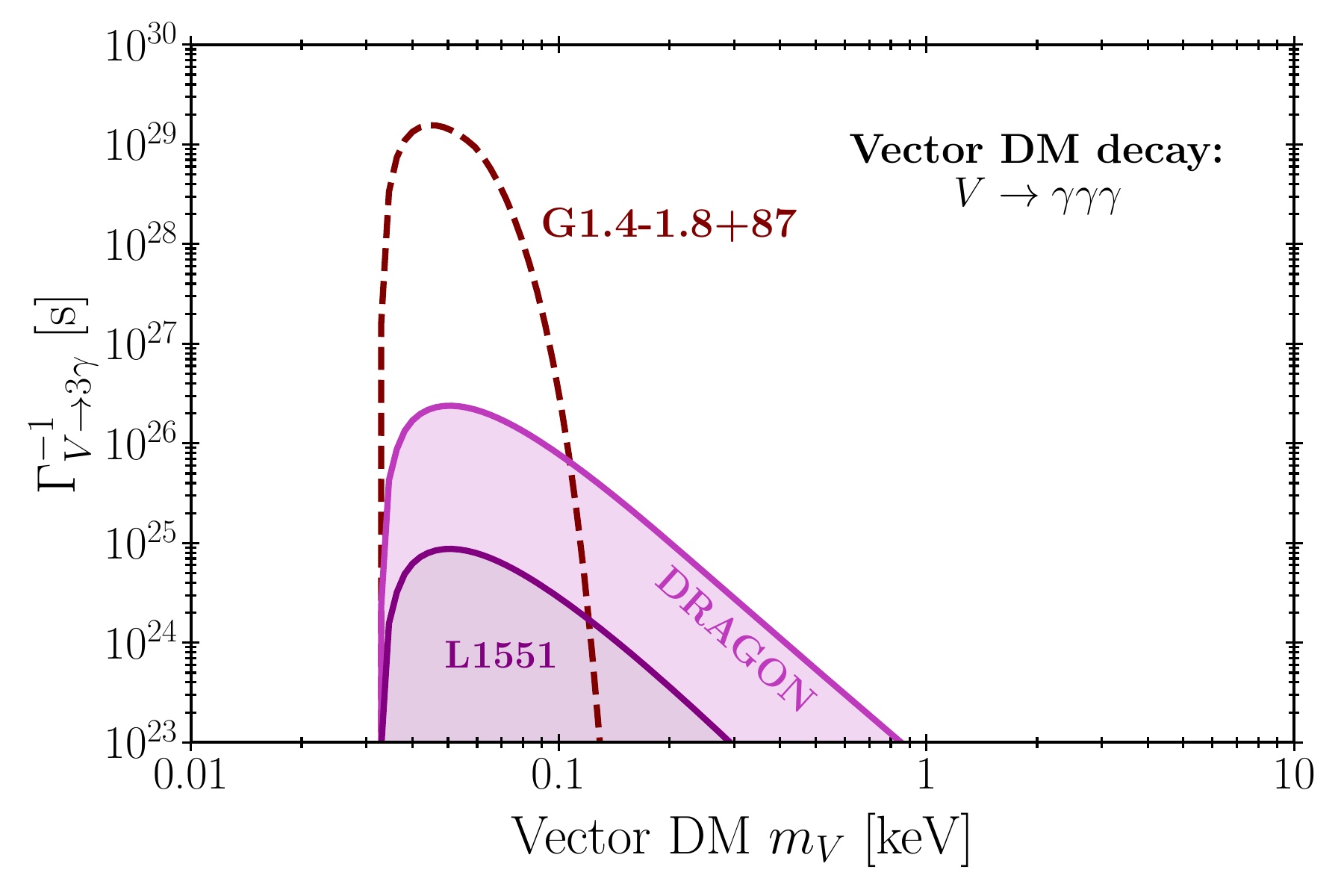}
\caption{\textbf{Left panel:} Product of DM-induced photon flux times ionization cross section, J$_{\chi}  \sigma_{H_2}$, for the decay of a vector particle into three photons. This serves as a direct indicator of the ionization rate, for several DM masses, and assumes $\rho_{\chi} =0.4$~GeV/cm$^{3}$. \textbf{Right panel:} Constraints on the lifetime of such a vector DM particle. The purple line represents constraints from {\it L1551}, the magenta line represents constraints from the {\it DRAGON} cloud and the dashed maroon line represents the optimistic forecast based on {\it G1.4-1.8+87}.}
\label{fig:dec3g}
\end{figure}

Finally, we have also calculated these constraints for the case of vector bosons decaying into three photons. This can be the case of dark photons or the B-L vector, which are popular candidates for DM particles. In this case, the spectrum of photons is given by a continuum spectrum, instead of a line-like spectrum.
The photon flux at a given position $J(E,\textbf{x})$ can be expressed as~as~\cite{Pospelov:2008jk, Linden:2024uph, McDermott:2017qcg, Cheung:2025gdn, Nguyen:2025eva, Redondo:2008ec, Krnjaic:2022wor, Linden:2024fby}
\begin{equation}
    \frac{dN_{V\rightarrow 3\gamma}}{dE_{\gamma}} = \frac{2}{E_{\gamma}} \frac{x^3}{51} \left(1715 - 3105\,x  + \frac{2919}{2}\, x^3 \right) 
     \,,
\label{eq:fluxTridents}
\end{equation}
where $x = 2E_{\gamma}/m_{V}$. Using this spectrum, we compute J$_{\chi \rightarrow3\gamma}$ for three-photon final state. The result is shown in Fig.~\ref{fig:dec3g}, upper panel. A trend similar to the monochromatic injection appears, with a larger ionization at lower DM masses. 
In Fig.~\ref{fig:dec3g} (lower panel) we show the limit on the decay lifetime obtained for this injection of particles, for the case of  {\it L1551} and {\it G1.4-1.8+87}.

We derive constraints on well-motivated vector dark matter models, such as the dark photon and the $B-L$ vector boson, using decay rate limits of the vector-to-three-photon final state. Both models arise from extensions of the Standard Model by anomaly-free $U(1)$ gauge symmetries. The dark photon interacts with Standard Model particles via kinetic mixing $\epsilon$ with the photon field~\cite{Holdom:1985ag, Fayet:1980ad, Fayet:1980rr, Kahn:2016vjr, Fabbrichesi:2020wbt, Caputo:2021eaa}, while the $B-L$ vector boson couples directly to the Standard Model through a flavor-universal coupling $g_{B-L}$~\cite{Chun:2022qcg, Carena:2004xs, Nguyen:2022zwb, Smith:2024jve}. These symmetry extensions are extensively studied in the contexts of string theory, string compactification, and vector dark matter production via quantum fluctuations during inflation~\cite{Servant:2002aq, Graham:2015rva, Hebecker:2023qwl, Cyncynates:2023zwj, Cyncynates:2024yxm, Carenza:2025uwx}.

%However, as we show in Fig.~\ref{fig:vectorConst} the resulting limits are not competitive with current direct detection constraints. In the left panel, we show our optimistic case in comparison with limits from XENON~\cite{}, INTEGRAL~\cite{} and those from imposing a decay rate lower than the age of the Universe, for the case of dark photons. The right panel reports a similar comparison, but for the case of B-L vector bosons, for which the decay into photons is largely suppressed and our bounds cannot be competitive.

\begin{figure}[h!]
\includegraphics[width=0.49\textwidth]{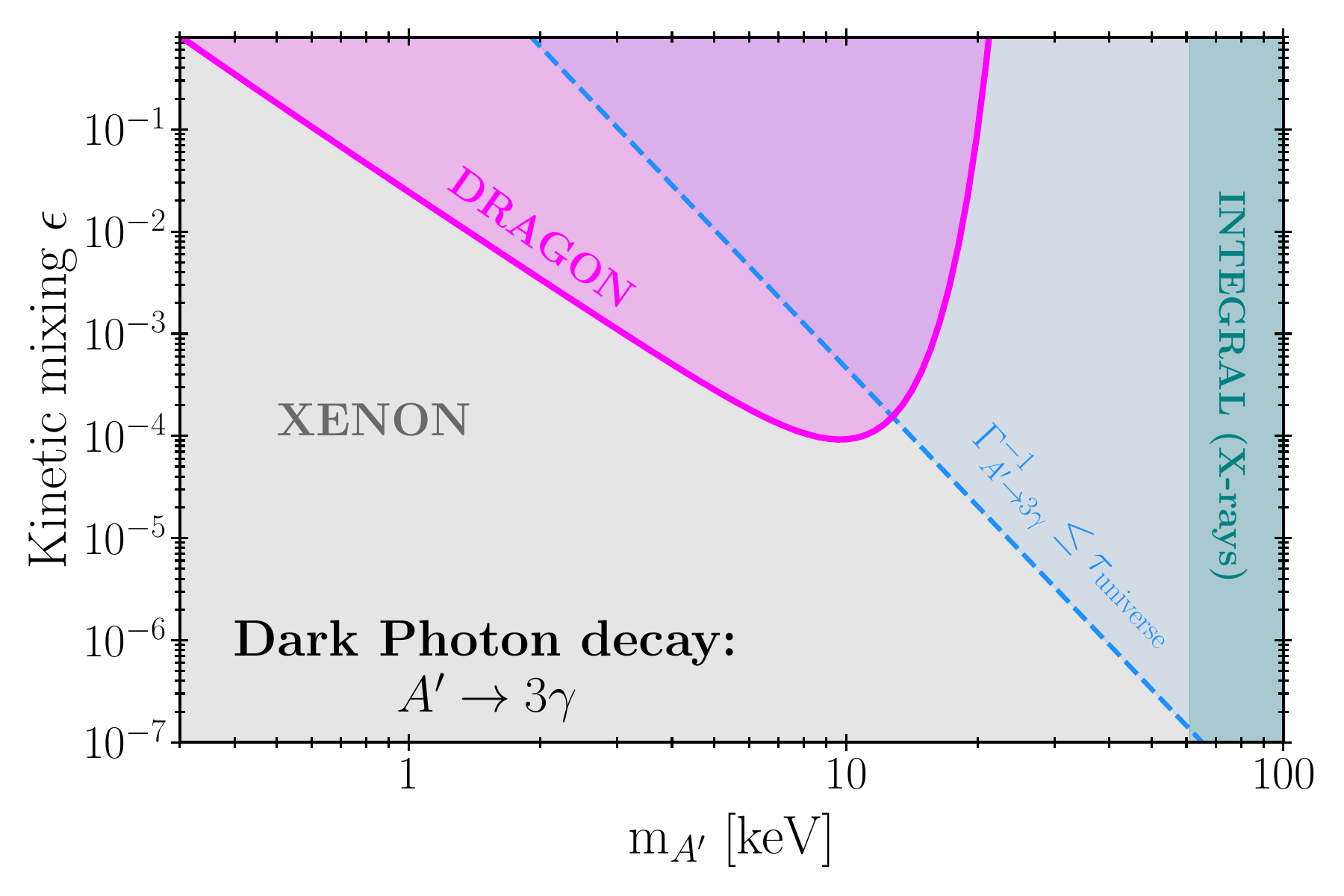}
\includegraphics[width=0.49\textwidth]{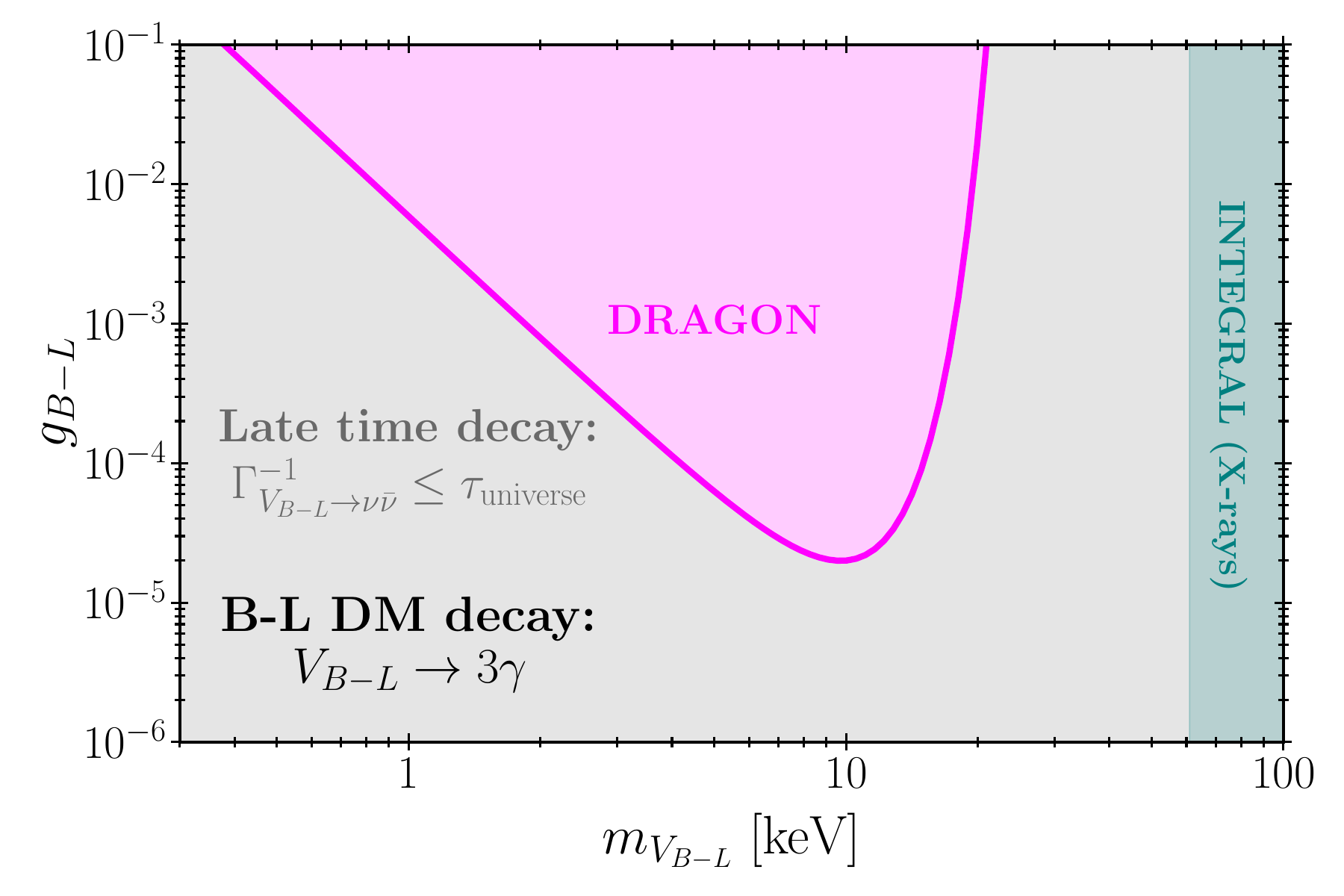}
\caption{Comparison of our optimistic forecast with existent constraints for the couplings of two popular vector boson DM candidates. \textbf{Left panel:} Bounds for the kinetic mixing of dark photons with limits from XENON~\cite{XENON:2019gfn} (gray), INTEGRAL~\cite{Linden:2024fby} (teal) and those from imposing a decay rate lower than the age of the Universe (dodgerblue). \textbf{Right panel:} Similar to the left panel but for the flavor-universal coupling $g_{B-L}$ of $B-L$ bosons with SM particles. The late time decay constraint assume $\nu\bar{\nu}$ final state is in gray. }
\label{fig:vectorConst}
\end{figure}

Figure~\ref{fig:vectorConst} presents the optimistic \emph{G1.4-1.8+87} constraints on vector dark matter models. We compute the decay width using the Euler-Heisenberg (EH) approximation, which is valid in the keV mass range~\cite{McDermott:2017qcg, Linden:2024fby}. The left panel shows the constraint on kinetic mixing $\epsilon$, compared to limits from the universe's age (4.6 Gyr), x-ray observations by INTEGRAL~\cite{Linden:2024fby}, and direct detection bounds from XENON~\cite{XENON:2019gfn}. For $B-L$ vector dark matter, we derive constraints on the flavor-universal coupling $g_{B-L}$, assuming the decay proceeds dominantly via electron loops under the EH approximation. We compare this result with the INTEGRAL limit on the same three-photon final state~\cite{Linden:2024fby} and with theoretical constraints based on neutrino decay channels and the universe's lifetime~\cite{Chun:2022qcg}.

For the dark photon, the \emph{G1.4-1.8+87} constraint surpasses the theoretical bound, as the three-photon decay dominates in this mass range. However, this region is already excluded by direct detection experiments that consider dark photon absorption. In contrast, our \( B\!-\!L \) constraints lie above the bound set by the universe's lifetime, assuming decay into neutrino pairs—the dominant channel for the \( B\!-\!L \) vector boson. This behavior extends to other lepton-flavor-dependent vector dark matter models such as \( L_i - L_j \). Overall, our results represent conservative limits on these classes of vector dark matter models.

\end{document}